\documentclass[sigconf,10pt]{acmart}

\renewcommand\footnotetextcopyrightpermission[1]{}

\usepackage{titlesec}
\usepackage{graphicx}
\usepackage{subcaption}
\titleformat*{\subsection}{\Large\bfseries}
\titleformat*{\subsection}{\large\bfseries}
\titleformat*{\subsubsection}{\large\bfseries}
\usepackage{tikz}
\usepackage{xcolor}
\usepackage{amsmath}

\usepackage{graphicx, epstopdf, stfloats, bbding, capt-of}

\usepackage{algorithm}
\usepackage{algorithmic}
\usepackage{enumitem }
\usepackage{multirow, multicol, booktabs, tabulary, tabu, longtable, array, varwidth}
\usepackage{placeins, lipsum}
\usepackage[flushleft]{threeparttable}
\usepackage{tablefootnote}
\usepackage{booktabs} %
\usepackage{makecell} %
\usepackage[bottom]{footmisc}

\newcommand{\parab}[1]{\vspace{0.5\baselineskip}\noindent\textbf{#1}~}

\newcommand{\ie}{\emph{i.e.,}\xspace}

\newcounter{BONumberOfComments}
\stepcounter{BONumberOfComments}

\newcounter{BeitongNumberOfComments}
\stepcounter{BeitongNumberOfComments}

\newcounter{LingzhiNumberOfComments}
\stepcounter{LingzhiNumberOfComments}

\newcounter{FYYNumberOfComments}
\stepcounter{FYYNumberOfComments}

\usepackage{cleveref}
\crefformat{section}{\S#2#1#3} %
\crefformat{subsection}{\S#2#1#3}
\crefformat{subsubsection}{\S#2#1#3}

\usepackage{xspace}
\newcommand{\sysname}{$\sf{AquaScope}$\xspace}

\makeatletter
\renewcommand{\@titlefont}{\LARGE\sffamily\bfseries}
\makeatother

\title{\sysname: Reliable Underwater Image Transmission on Mobile Devices}

\newcommand{\colspace}{\hspace{20pt}}

\author{
  \Large
  \begin{tabular}{c @{\colspace} c @{\colspace} c @{\colspace} c}
    Beitong Tian$^*$ & Lingzhi Zhao$^*$ & Bo Chen & Mingyuan Wu \\
    Haozhen Zheng & Deepak Vasisht & Francis Y. Yan & Klara Nahrstedt
  \end{tabular}
}
\renewcommand{\authors}{Beitong Tian, Lingzhi Zhao, Bo Chen, Mingyuan Wu,
Haozhen Zheng, Deepak Vasisht, Francis Y. Yan, Klara Nahrstedt}
\renewcommand{\shortauthors}{B. Tian, L. Zhao, B. Chen, M. Wu,
H. Zheng, D. Vasisht, F. Y. Yan, K. Nahrstedt}

\affiliation{
  \vspace{8pt}
  \institution{\Large University of Illinois Urbana-Champaign}
  \country{}
  \vspace{10pt}
}

\thanks{$^*$Both authors contributed equally to this work.}

\begin{document}

\begin{abstract}
Underwater communication is essential for both recreational and
scientific activities, such as scuba diving.
However, existing methods remain highly constrained by environmental
challenges and often require specialized hardware,
driving research into more accessible
underwater communication solutions.
While recent acoustic-based communication systems support
text messaging on mobile devices, their low data rates
severely limit broader applications.

We present \sysname, the first acoustic communication system
capable of underwater image transmission on commodity mobile devices.
To address the key challenges of underwater environments---limited
bandwidth and high transmission errors---\sysname employs and enhances
generative image compression to improve compression efficiency,
and integrates it with reliability-enhancement techniques at the
physical layer to strengthen error resilience.
We implemented \sysname on the Android platform and demonstrated its
feasibility for underwater image transmission.
Experimental results show that \sysname enables reliable, low-latency
image transmission while preserving perceptual image quality,
across various bandwidth-constrained and error-prone underwater conditions.
\end{abstract}

\maketitle
\pagestyle{plain}

\section{Introduction}\label{sec:introduction}

\begin{figure}[t]
\centering
\includegraphics[width=\linewidth]{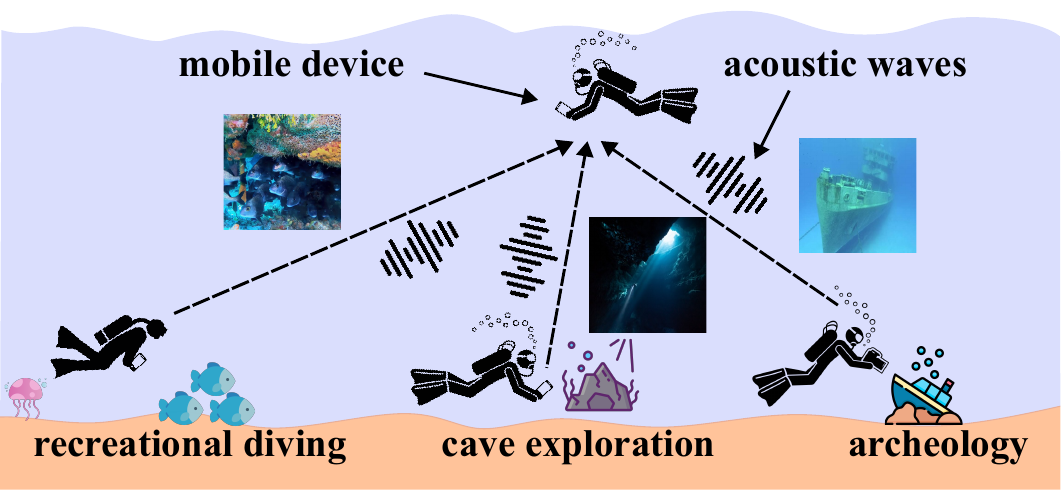}
\caption{Conceptual graph of an underwater image transmission system for scuba divers.}
\label{fig:conceptual_graph}
\end{figure}

Underwater activities, e.g., snorkeling and scuba diving,
are among the primary ways humans explore the underwater world,
with tens of millions participating annually for recreational or scientific
purposes~\cite{stat_zhao_plus,stat_zhao,stat_zhao_1}.
Effective underwater communication is crucial for people to share experiences,
follow instructions, and alert others to dangers.

However, underwater communication has historically been limited due to
fundamentally challenging underwater environments
(e.g., scuba divers still largely rely on hand signals to communicate today).
Wireless technologies, such as cellular, Wi-Fi, or Bluetooth,
are ineffective underwater because radio frequency (RF) signals
suffer substantial attenuation in aquatic
environments~\cite{chen2022underwater,chen2023underwater}.
While wired networks offer an alternative, they are often too cumbersome and
impractical underwater.
Existing research~\cite{afzal2022battery,tonolini2018networking,jang2019underwater,carver2021amphilight,mobicom24_seascan} addressing these limitations
relies on specialized, sophisticated hardware, which is not accessible for
everyday use.

The latest advancement in ubiquitous underwater communication
utilizes acoustic signals to enable text messaging on mobile
devices~\cite{chen2022underwater}, as acoustic signals are known to propagate
better in water than RF signals. Nevertheless, with its limited data rate---only 0.6 kbps beyond 10 meters---its practical applications are highly restricted.

In this study, we investigate the feasibility of \emph{underwater
image transmission}. As envisioned in Fig.~\ref{fig:conceptual_graph},
enabling image transmission would enhance communication
between scuba divers and enrich the overall experience of various
underwater activities.
Despite active research efforts in both industry and
academia~\cite{gofish, AppleRD2021, liu2023efficient,afzal2022battery},
implementing this technology faces two major challenges
posed by underwater environments: (1) severely limited bandwidth, which renders
conventional image codecs infeasible due to their inadequate compression
efficiency, and (2) high transmission errors (caused by varying channel
conditions), which may substantially degrade
reconstructed image quality or even prevent image decoding altogether.

We present \sysname, the first underwater communication system
that enables image transmission on commodity mobile devices.
\sysname is also acoustic-based, but it is specifically designed
to optimize bandwidth efficiency and enhance error resilience.
At its core, \sysname employs \textit{generative image
compression}~\cite{mao2024extreme,yu2024image,santurkar2018generative},
a recent technique that achieves significantly higher compression rates
than traditional image codecs (e.g., JPEG~\cite{wallace1991jpeg})
and neural codecs~\cite{cheng2020image}.
This generative model encodes an image into a sequence of
``tokens,'' each representing a small patch (e.g., 16$\times$16 pixels) of the
original image in a more bandwidth-efficient manner
(\S\ref{sec:generative_compression}).
These tokens are then modulated by \sysname into acoustic signals
and transmitted underwater.
If a token is corrupted or lost during transmission,
the image decoder can leverage its ``generative'' capabilities
to reconstruct the image, thereby mitigating transmission errors.

While the above design appears intuitive, its practical effectiveness
is hindered by several challenges.
First, generative compression models are typically pre-trained on large
image datasets spanning diverse scenes. However, in the domain-specific
underwater scenario, our analysis (Fig.~\ref{fig:codebook_merger}) reveals
that more than half of the tokens in the codebook are never used,
leading to wasted bandwidth.
Second, existing generative models assume that all encoder-generated tokens
are received intact by the decoder. In reality, underwater transmission
is prone to errors that can distort tokens.
Third, the constrained computational and communication resources on commodity mobile devices
necessitate efficient reliability-enhancing techniques at the physical (PHY)
layer to correct errors in data symbols and ensure that tokens
remain recoverable by the generative decoder after demodulation.
We address these challenges using three key techniques.

\parab{Context-aware tokens distillation (\S\ref{sec:semantic_filtering}).}
To reduce the number of tokens used by a generic generative
compression model to encode an image, we employ a ``distillation'' process
that transfers the most relevant information from the original token set
to a significantly smaller set, motivated by the fact that the original
tokens are heavily underutilized in underwater scenarios.
Our distillation process involves training a transformer model in context,
on underwater image datasets, enabling a higher compression ratio
while preserving perceptual image quality.

\parab{Error-resilient fine-tuning (\S\ref{sec:fine_tuning}).}
To enhance the error resilience of a generative compression model,
we fine-tune its decoder on underwater datasets
under simulated random token disturbances.
While this approach resembles prior loss-resilient codecs
proposed for video
systems~\cite{nsdi24_grace,li2024reparolossresilientgenerativecodec},
a fundamental difference in \sysname is that underwater transmission errors
distort data symbols directly at the PHY layer,
making it impossible for the decoder to explicitly identify and mask
corrupted (or lost) tokens, as done in prior methods.
As a result, we must strategically perturb tokens during training,
ensuring that the decoder learns to reconstruct images without prior knowledge
of corrupted tokens.

\parab{Reliability enhancement at the PHY (\S\ref{sec:packetizer},
\S\ref{sec:synchronization}).}
Improving error resilience at the PHY layer is equally important,
as it minimizes the number of corrupted tokens passed to the generative
image decoder after demodulation and helps preserve reconstructed
image quality.
To achieve this, we first insert multiple training symbols into
the data packet. These training symbols enable the receiver to perform
time equalization for channel estimation, improving the recovery of data
symbols. Next, to address symbol time offsets caused by underwater device
movement, we propose an effective time synchronization algorithm
based on two key observations: the relative speed between the sender and
receiver is inherently limited underwater,
and their distance changes gradually without sudden shifts.
Leveraging these physical principles,
\sysname filters out unrealistic detected start timestamps and smooths detection
results, thereby obtaining accurate start timestamps for symbols.

We developed a prototype of \sysname on the Android platform
and conducted extensive real-world experiments.
Our results show that:
(1) \sysname reliably transmits 256$\times$256 color images with high fidelity
in under 9s over distances of up to 20m;
(2) \sysname maintains robust performance across various depths,
orientations, distances, and mobility conditions, reducing the bit error
rate (BER) from 19\% to below 2\% on average compared with state-of-the-art
systems;
(3) \sysname achieves an average LPIPS (Learned Perceptual Image Patch
Similarity)~\cite{zhang2018unreasonable} score of 0.3, indicating strong
visual consistency between transmitted and receive images.

Our key contributions are as follows:
\begin{itemize}
[topsep=2pt,noitemsep,leftmargin=*]
\item We design and implement \sysname, the first underwater acoustic system
that enables reliable image transmission between mobile devices.
\item \sysname employs generative image compression and enhances its
bandwidth efficiency and error resilience for underwater communication
(through context-aware token distillation and error-resilient fine-tuning).
\item \sysname implements reliability-enhancing techniques at the PHY
layer to mitigate the impact of underwater transmission errors.
\item \sysname demonstrates the feasibility of image transmission in
bandwidth-limited, error-prone underwater environments on
resource-constrained mobile devices.
\end{itemize}

\section{Motivation}\label{sec:background}

Existing underwater mobile communications rely on acoustic signals,
as they propagate more effectively in water than RF signals (e.g., Wi-Fi and Bluetooth).
In this section, we characterize the harsh channel conditions encountered by
mobile devices underwater and demonstrate the limitations of a state-of-the-art
underwater acoustic system, AquaApp~\cite{chen2022underwater},
in supporting image transmission. These challenges underscore the need for an
alternative solution.

\subsection{Characterizing Mobile Devices Underwater}\label{sec:mov_channel}

\begin{figure*}[t]
    \centering
    \begin{minipage}[b]{0.24\textwidth}
        \centering
        \includegraphics[width=\textwidth]{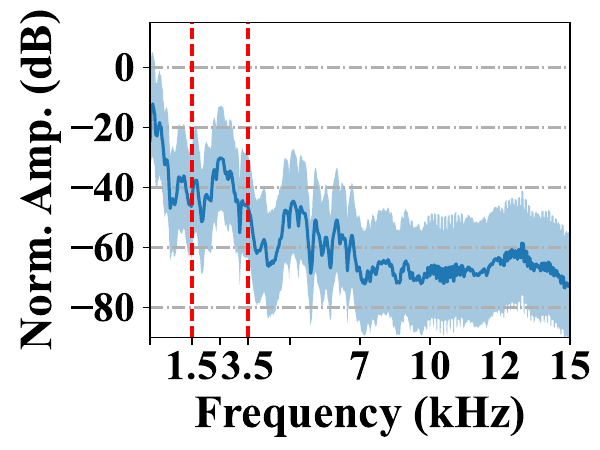}
        \caption{Channel response.}
        \label{fig:snr}
    \end{minipage}
    \begin{minipage}[b]{0.24\textwidth}
        \centering
        \includegraphics[width=\textwidth]{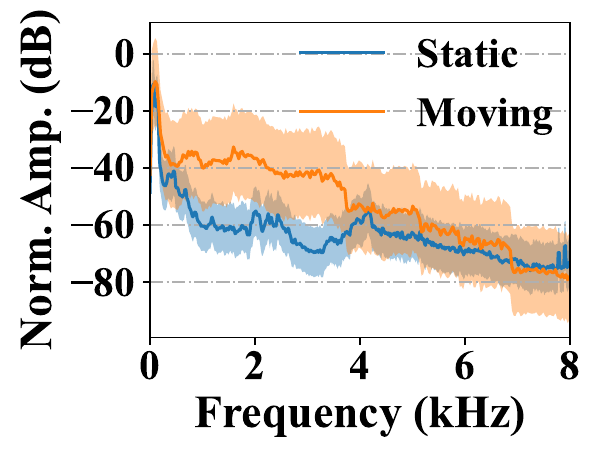}
        \caption{Noise.}
        \label{fig:noise}
    \end{minipage}
    \begin{minipage}[b]{0.24\textwidth}
        \centering
        \includegraphics[width=\textwidth]{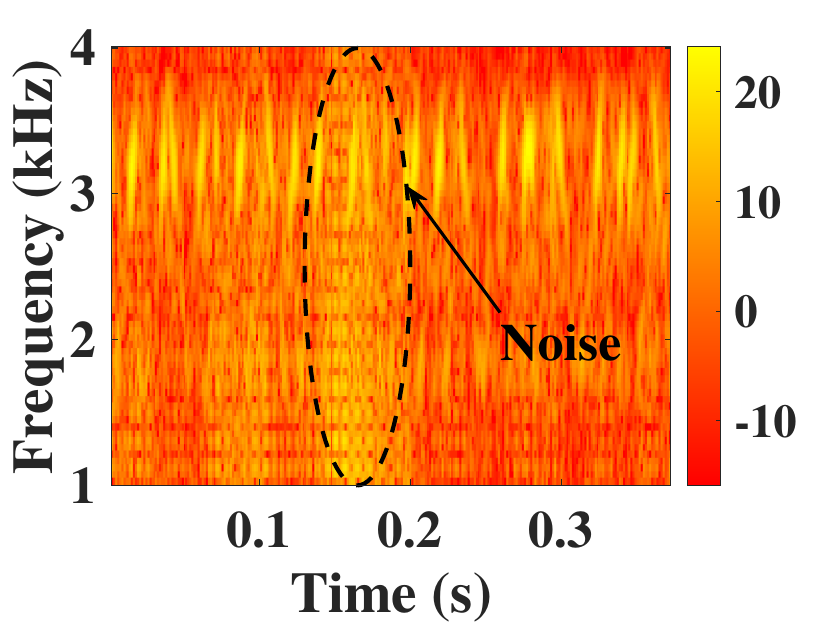}
        \caption{Spectrogram.}
        \label{fig:spectrogram}
    \end{minipage}
    \begin{minipage}[b]{0.24\textwidth}
        \centering
        \includegraphics[width=\textwidth]{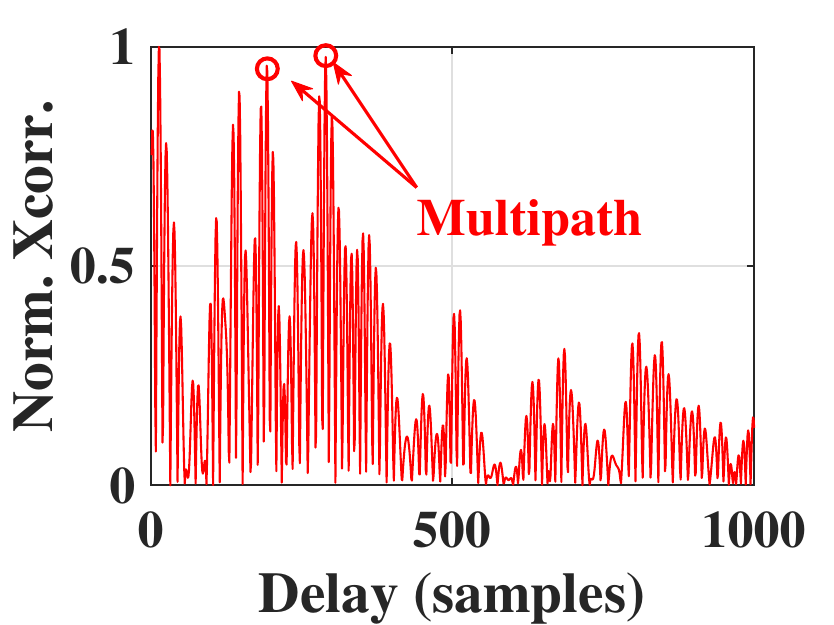}
        \caption{Cross-correlation.}
        \label{fig:multipath}
    \end{minipage}
\end{figure*}

\noindent\textbf{Frequency selectivity.} The frequency range of audible signals transmitted and received by a mobile phone's speaker and microphone is inherently limited and further influenced by the frequency-selective nature of the underwater channel. To illustrate this effect, we submerge two commercial Samsung S21~\cite{Samsung} phones in an open lake at a depth of 2m and position them 5m apart. One phone transmits 0--15 kHz chirp signals,
each lasting 1s.
Fig.~\ref{fig:snr} shows the frequency response of the underwater channel in terms of normalized amplitude (dB), with the shaded area representing variations.
The results indicate a significant drop in frequency response at 3.5 kHz,
suggesting that acoustic signals beyond this threshold experience
rapid attenuation~\cite{chen2022underwater,yang2023aquahelper}.

\parab{Ambient noise.} Underwater ambient noise, such as air bubbles, water flow, and sounds produced by the waterproof bag, overlaps with the frequency bands of a mobile phone's acoustic signal, leading to additional interference.
We use a Samsung S21 phone to record underwater ambient noise in the same lake
under both static and moving conditions.
Fig.~\ref{fig:noise} presents the normalized amplitude of ambient noise across the
0--8 kHz range, showing particularly high noise levels between 0--1 kHz.
Moreover, ambient noise increases significantly when the device is in motion,
primarily due to rustling and friction-related noises generated by the waterproof
bag.
Spectrum analysis in Fig.~\ref{fig:spectrogram} further indicates that this noise
spans frequencies from 1--4kHz, posing considerable challenges to acoustic
signal transmission.

\parab{Multipath.} The underwater environment introduces significant multipath effects,
as acoustic waves are reflected by the lake's surface, bottom, rough rocks, and other
objects. These reflections result in both inter- and intra-symbol interference.
Fig.~\ref{fig:multipath} illustrates this phenomenon, showing multiple delayed replicas
of the main signal.

\subsection{Underwater Image Transmission Challenges}\label{sec:mov_challenges}

Faced with substantial challenges posed by underwater
environments (\S\ref{sec:mov_channel}), the state-of-the-art acoustic
communication system---AquaApp~\cite{chen2022underwater}---only supports text
messaging on mobile devices.
In this study, we push the boundaries of underwater communication by
exploring the feasibility of image transmission.

We implement AquaApp on Samsung S21 mobile phones and extend it to support
image transmission. Specifically, images are encoded into byte streams
using different codecs, including
JPEG~\cite{wallace1991jpeg}, PNG~\cite{PNG}, and recent neural
codecs~\cite{cheng2020image}. These encoded bytes are then transmitted
using AquaApp's three-step OFDM (Orthogonal Frequency Division Multiplexing)
protocol. The process begins with the sender transmitting a preamble to the
receiver. Based on this preamble, the receiver identifies and replies with high
SNR (signal-to-noise ratio) subcarriers.
The sender then encodes the payload on the selected subcarriers and transmits
the data.

\begin{figure}[t]
    \centering
    \begin{minipage}[b]{0.23\textwidth}
        \centering
        \includegraphics[width=\textwidth]{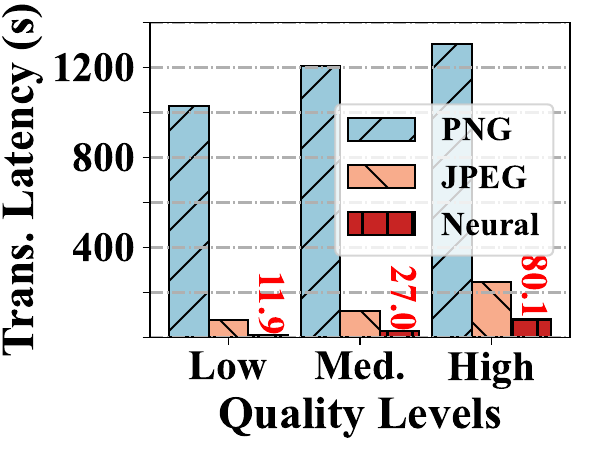}
        \caption{Transmission laten\-cy of images compressed with different codecs.}
        \label{fig:BPP}
    \end{minipage}\hfill
    \begin{minipage}[b]{0.23\textwidth}
        \centering
        \includegraphics[width=\textwidth]{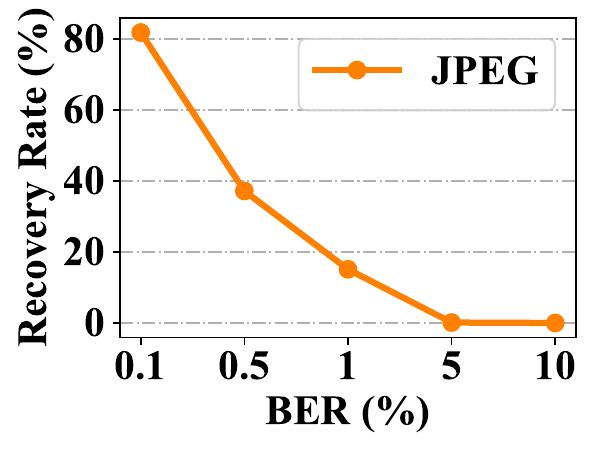}
        \caption{Recovery rate (PNG and neural codecs fail to reconstruct images).}
        \label{fig:recovery}
    \end{minipage}
\end{figure}

\begin{figure}[t]
    \centering
    \begin{subfigure}[b]{.22\linewidth}
        \centering
        \includegraphics[width=\linewidth]{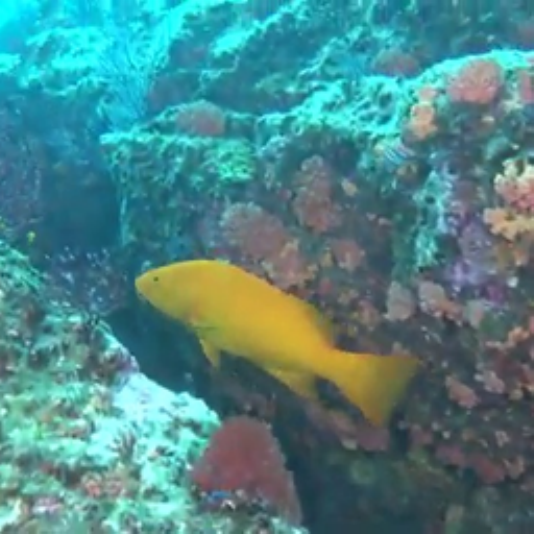}
        \subcaption{Raw}
        \label{fig:00009_frame14_origin}
    \end{subfigure}
    \begin{subfigure}[b]{.22\linewidth}
        \centering
        \includegraphics[width=\linewidth]{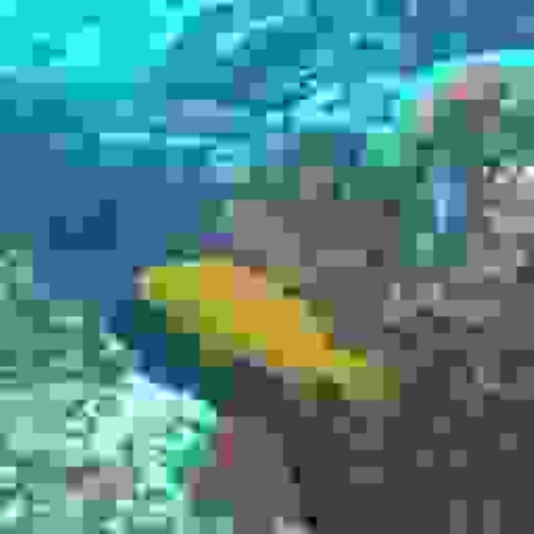}
        \subcaption{JPEG}
        \label{fig:00009_frame14_origin_quality_1}
    \end{subfigure}
    \begin{subfigure}[b]{.22\linewidth}
        \centering
        \includegraphics[width=\linewidth]{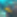}
        \subcaption{PNG}
        \label{fig:00009_frame14_origin_8x8}
    \end{subfigure}
        \begin{subfigure}[b]{.22\linewidth}
        \centering
        \includegraphics[width=\linewidth]{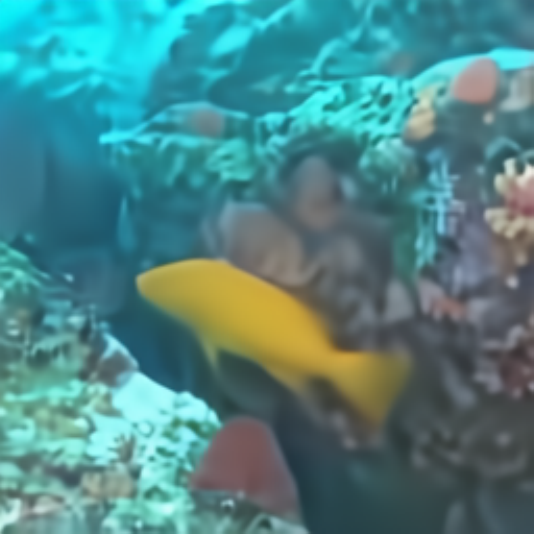}
        \subcaption{NN}
        \label{fig:00009_frame14_origin_8x8}
    \end{subfigure}
    \caption{Visual comparison of different image codecs at their lowest quality settings.}
    \label{fig:semantic_vs_pixel}
\end{figure}

\parab{Image codecs.} First, we show that existing codecs exhibit \textit{low compression efficiency} and are \textit{vulnerable to transmission errors}.
Fig.~\ref{fig:BPP} illustrates the transmission latency for a 256$\times$256 image at various compression levels.
Since AquaApp's maximum effective data rate is only 0.6 kbps~\cite{chen2022underwater}, it takes at least 80s, 1000s, 12s to transmit images
encoded with JPEG (low quality level of 30), PNG (scaled to 30\% of the original), and neural codecs (low quality level of 1), respectively, \textit{assuming} no transmission errors. Moreover, low-quality settings introduce noticeable artifacts and blurring, as shown in Fig.~\ref{fig:semantic_vs_pixel}, significantly reducing image clarity. This is because these codecs prioritize
the reconstruction of pixel-level details, which proves inefficient for underwater
transmission.

Meanwhile,
errors are inevitable over extended transmission times due to channel variations. Existing image codecs are particularly vulnerable to transmission errors due to entropy coding~\cite{huffman1952method,witten1987arithmetic}.
If a received packet is corrupted or incomplete, the image decoding process
may fail.
To quantify this, we define the \textit{recovery rate} as the
fraction of successfully decoded images among transmitted ones.
We simulate lossy channels with varying bit error rates (BERs) and compute
the recovery rate over 500 images.
The results, shown in Fig.~\ref{fig:recovery}, indicate that more than half
of JPEG images fail to decode when the BER exceeds 0.5\%.
PNG- and neural network-compressed images are unrecoverable even at a BER
of 0.1\%, so their results are omitted.

\begin{figure}[t]
    \centering
    \begin{minipage}{0.23\textwidth}
        \centering
        \includegraphics[width=\textwidth]{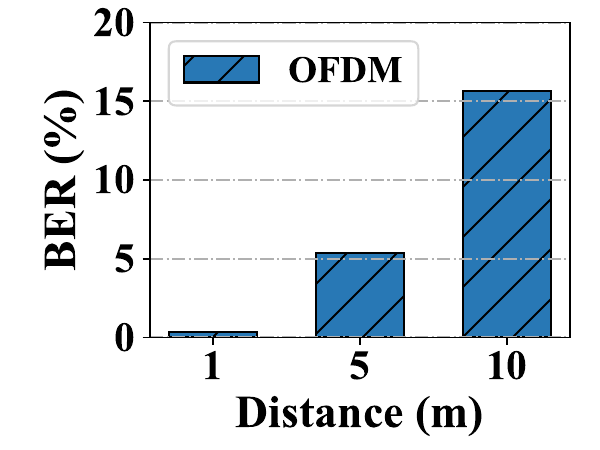}
        \caption{BER {\em vs.} distance.}
        \label{fig:limitation_ofdm_distance}
    \end{minipage}\hfill
    \begin{minipage}{0.23\textwidth}
        \centering
        \includegraphics[width=\textwidth]{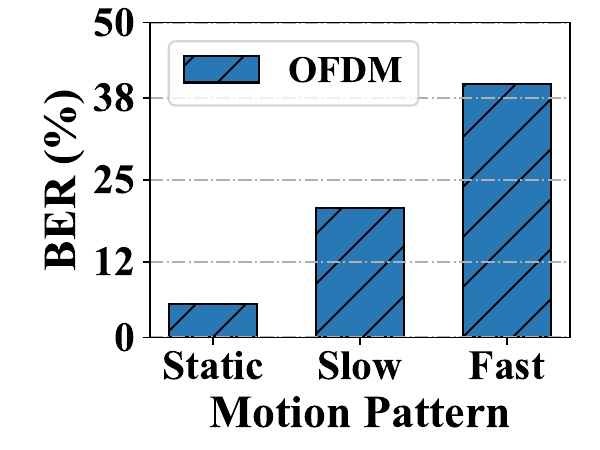}
        \caption{BER {\em vs.} mobility.}
        \label{fig:limitation_ofdm_mobility}
    \end{minipage}
\end{figure}

\parab{Transmission protocol.} Second, we demonstrate the limitations of the OFDM protocol employed by AquaApp.
Fig.~\ref{fig:limitation_ofdm_distance} and Fig.~\ref{fig:limitation_ofdm_mobility} illustrate the BER under varying distance and mobility conditions, respectively.
Even in the least demanding scenario, we observe that the BER reaches 0.36\%,
corresponding to a recovery rate of only about 60\% (Fig.~\ref{fig:recovery}).
Furthermore, the BER increases rapidly when the distance exceeds 5 meters or
when the device is in motion, reaching up to 40\% and rendering underwater image transmission infeasible.

Since AquaApp is designed for small packet transmissions (\ie text messages), its transmission protocol struggles with channel dynamics, extended transmission ranges, and mobility. The decline in SNR with increasing transmission distance leads to high demodulation errors, even with carefully selected subcarriers.
Moreover, device movement can alter channel conditions and cause previously
selected subcarriers to degrade in quality over the course of transmission.
Addressing this requires frequent subcarrier reselection, which is
time-consuming (e.g., $\sim$3 seconds per selection).

\subsection{New Opportunity: Generative Compression}\label{sec:generative_compression}

Given the harsh conditions of underwater communication, an effective image
compression scheme must reduce the image size substantially while achieving
robustness against transmission errors, posing a challenge for existing image
codecs~\cite{wallace1991jpeg,webp2024,cheng2020image}. However, recent
advancements in deep generative
models~\cite{NIPS2014_5ca3e9b1,yu2024image,mao2024extreme} offer a promising
alternative---\textit{generative compression}.

\begin{figure}[t]
    \centering
    \includegraphics[width=\linewidth]{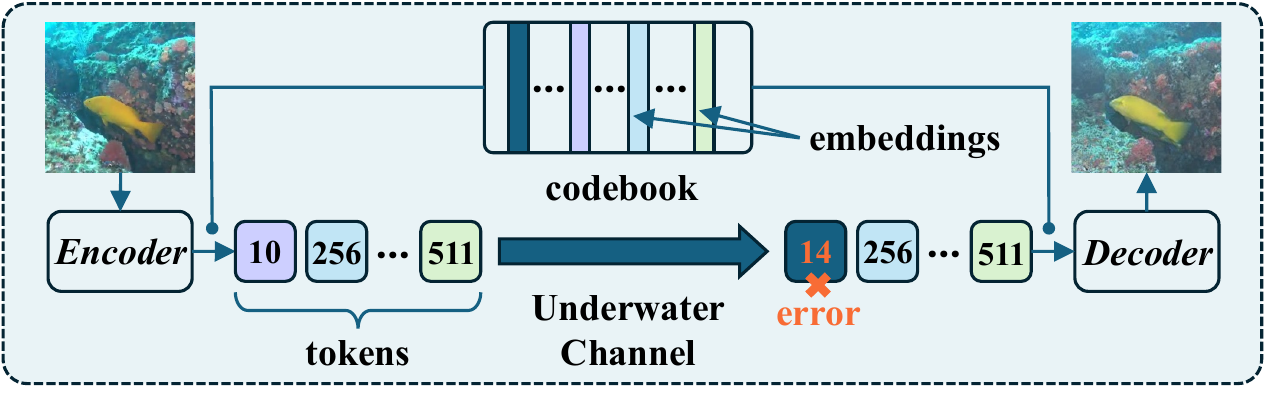}
    \caption{Overview of generative image compression.}
    \label{fig:semantic_compression}
\end{figure}

Fig.~\ref{fig:semantic_compression} depicts the workflow of generative compression.
The encoder transforms a raw image into a sequence of \emph{tokens},
which represent \emph{indices}\footnote{The terms ``tokens'' and ``indices'' are
commonly used interchangeably.} of embeddings (\ie feature vectors)
in a pretrained codebook.
Unlike previous neural codecs~\cite{cheng2020image} that encode images
as a large set of feature values, this token-based method provides
a more bandwidth-efficient representation of semantic information~\cite{yu2024image,mao2024extreme}.
After the image tokens are transmitted over the underwater channel,
the decoder maps them to their corresponding embeddings and
reconstructs the original image.

Generative compression achieves efficient compression with
``graceful degradation''~\cite{santurkar2018generative},
where reconstructed image quality gradually declines
as bandwidth decreases or transmission errors occur.
As we will show in our experiments, generative compression operates
at bandwidth levels beyond the reach of convention methods and reduces
transmission latency by half compared with neural codecs.
Moreover, even in the presence of transmission errors,
the received image remains both decodable and semantically meaningful.

\parab{Summary:}
\begin{itemize}[topsep=2pt,noitemsep,leftmargin=*]
   \item Existing underwater communication systems, designed for small packet transmissions, are inadequate for handling image data.%
   \item Both traditional and neural image codecs face significant challenges in underwater environments due to limited bandwidth and vulnerability to transmission errors.
   \item Generative image compression presents a promising solution for reliable underwater image transmission.
\end{itemize}

\section{System Overview}
\label{sec:system_overview}
\begin{figure}[t]
    \centering
    \includegraphics[width=\linewidth]{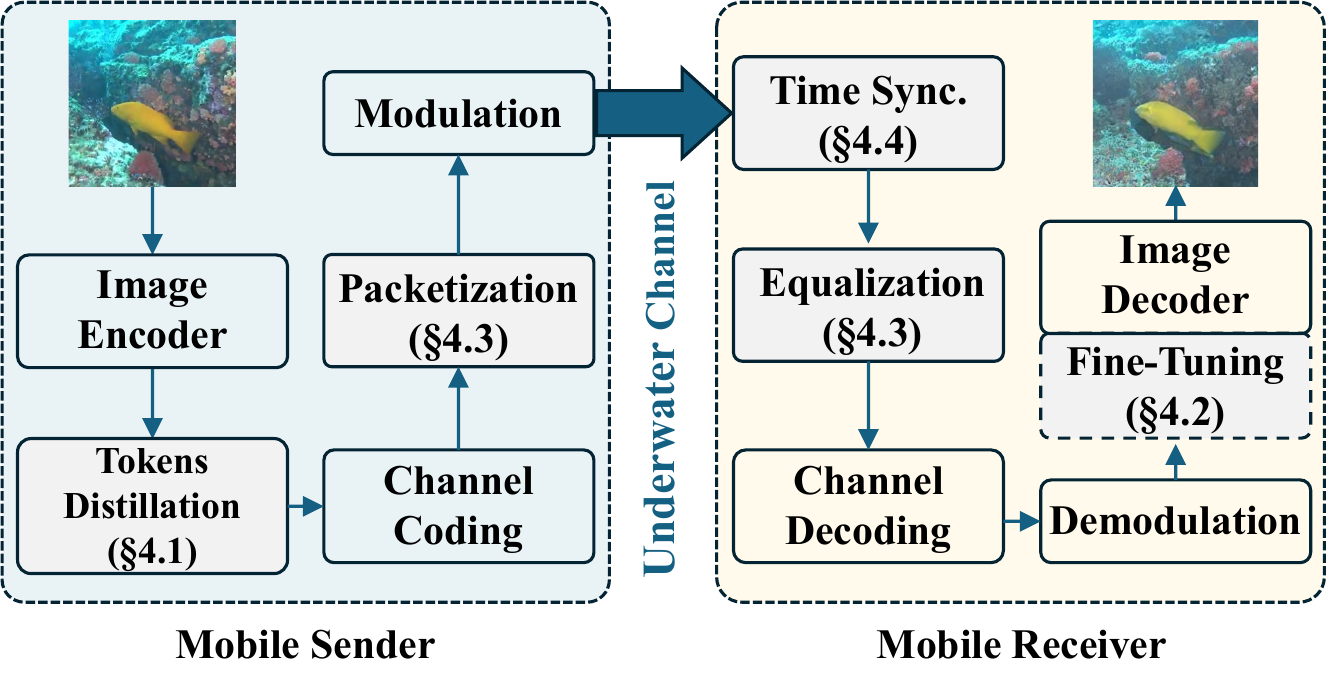}
    \caption{\sysname overview.}
    \label{fig:system_overview}
\end{figure}

Fig.~\ref{fig:system_overview} provides an overview of \sysname.
The image encoder and decoder are jointly trained offline on underwater datasets with simulated transmission errors (\cref{sec:fine_tuning}).
During operation, the mobile sender compresses
a raw image into tokens (i.e., indices of embeddings in a codebook)
and converts them into a more compact bitstream
using ``context-aware distillation'' (\cref{sec:semantic_filtering}).
It then encodes the bits into data symbols via channel coding and
encapsulates them into a packet at the PHY layer (\cref{sec:packetizer}).
This packet includes a leading preamble followed by multiple symbol groups.
Next, the sender modulates the packet data into an acoustic signal
for transmission through the underwater channel to the receiver.

Upon reception, the mobile receiver performs preamble detection and time synchronization (\cref{sec:synchronization}) to locate symbol groups in the received signal.
It then compensates for channel distortions through equalization (\cref{sec:packetizer}),
demodulates the signal into data symbols,
reconstructs the bitstream,
and ultimately generates a high-fidelity underwater image.

\section{\sysname Design}
\label{sec:systemdesign_transmission}
In this section, we introduce three key designs of \sysname:
context-aware tokens distillation, error-resilient fine-tuning, reliability enhancement at the PHY layer including packetization
and smoothed \& bounded time synchronization.

\subsection{Context-aware Tokens Distillation}
\label{sec:semantic_filtering}

Generative image compression models encode an image with $M$ tokens,
each selected from a codebook containing $K$ possible tokens.
This results in a compressed image size of $M\times \log_2 K$ bits.
Designed to reconstruct and generate images with high fidelity across
a wide range of scenes (e.g., ImageNet-1000 with 1,000 object classes),
these generative models often conservatively adopt large values for $M$ and $K$
to preserve visual details.
For example, VQGAN~\cite{mao2024extreme} uses $M=256$ and $K=1024$ for an input
resolution of 256$\times$256, resulting a compressed image size of 2,560 bits.

\begin{figure}[t]
    \centering
    \includegraphics[width=0.3\textwidth]{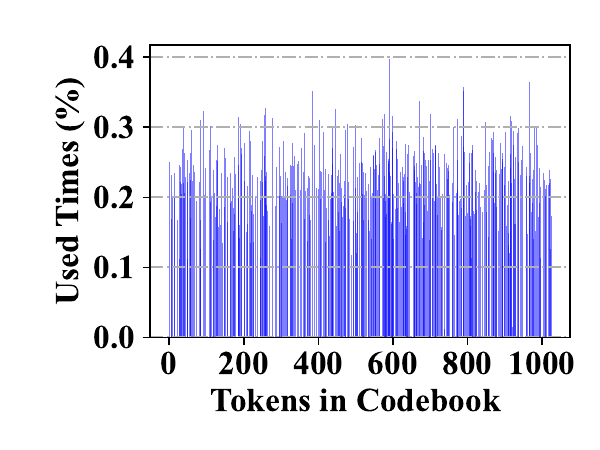}
    \vspace{-5pt}
    \caption{Tokens in codebook are underutilized: over 50\% of tokens remain unused in underwater image compression.}
    \label{fig:codebook_merger}
\end{figure}

However, in specialized domains such as underwater imaging, we observe
that many tokens in the codebook are rarely or never used, resulting in
inefficient bandwidth usage.
To quantify this redundancy, we analyze token utilization when compressing
images from underwater datasets.
As shown in Fig.~\ref{fig:codebook_merger}, 519 out of 1024 tokens in VQGAN's
codebook remain unused, presenting an opportunity to optimize
the representation and enhance compression efficiency.

Following the methodology laid out by TiTok~\cite{yu2024image},
a state-of-the-art generative model for image compression,
we employ a transformer-based distillation approach to improve token utilization
and minimize bandwidth consumption in underwater communication.
Specifically, given $M$ original image tokens, we append $M'$ randomly initialized
tokens ($M' \ll M$) and train a transformer model on underwater image datasets
to transfer the most relevant information from the original tokens to the
appended tokens (via the attention
mechanism~\cite{touvron2021training}).
On the decoder side, a similar transformer is trained to reconstruct the
original $M$ tokens from the received $M'$ tokens, by minimizing the
cross-entropy loss between the original and reconstructed tokens.
Through this \emph{context-aware tokens distillation} process, the data size
is reduced by a factor of $(M \times \log_{2}K) / (M'\times \log_{2}K')$, where $K'$
represents the codebook size for the distilled tokens.

\subsection{Error-resilient Fine-tuning}\label{sec:fine_tuning}

Generative image compression models are commonly pre-trained on large-scale
image datasets (e.g., ImageNet) to achieve generalization, with the training
process assuming that all tokens produced by the encoder are received intact
by the decoder.
However, in underwater environments, transmission errors are inevitable,
often resulting in corrupted or lost tokens.
Meanwhile, best practices suggest that fine-tuning a pre-trained generative
model for a specific target scenario enhances performance.
Therefore, we fine-tune the image decoder on underwater
datasets with simulated transmission errors, as detailed below.

\begin{figure}[t]
    \centering
    \includegraphics[width=\linewidth]{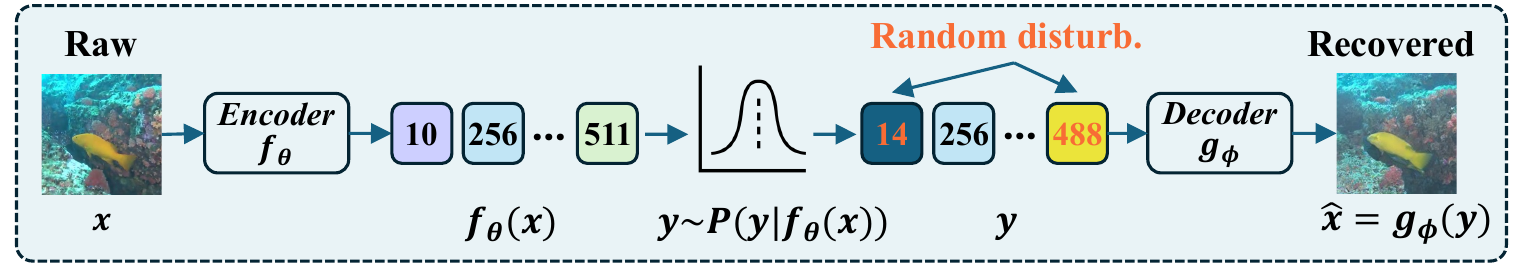}
    \caption{Fine-tuning generative image compression for underwater communication.}
    \label{fig:fine_tuning}
\end{figure}

We denote the encoder and decoder by $f_{\theta}$ and $g_{\phi}$, respectively,
with $\theta$ and $\phi$ representing their corresponding weights.
As illustrated in Fig.~\ref{fig:fine_tuning}, the encoder $f_{\theta}$
converts a raw image $\mathbf{x}$ into a sequence of indices
$f_{\theta}(\mathbf{x})$.
Then, we randomly sample from these indices and introduce disturbance.
The resulting sequence $\mathbf{y}$ is used by the decoder to
reconstruct the image as $\hat{\mathbf{x}} = g_{\phi}(\mathbf{y})$.
To enhance reconstruction quality, we optimize a composite loss function that
combines pixel-level reconstruction loss, perceptual loss, and GAN loss,
following standard practices in image reconstruction model
training~\cite{esser2021taming}.

The random disturbance is introduced as follows.
Given $M$ tokens, we randomly perturb $m$ of them (by replacing each
with an alternative token from the codebook), where $m$ is randomly selected
from $[1, p]$ ($p \le M$).
Following the curriculum learning strategy~\cite{bengio2009curriculum}, we progressively increase $p$ during training, allowing the model to learn with increasingly challenging perturbations for improved performance. We cap \( p \) at 25\% of \( M \) to prevent excessive perturbation, which could eliminate the useful information in tokens and hinder the model's ability to learn. This 25\% threshold is based on preliminary real-world experiments on error rates.

A key difference between \sysname and existing loss-resilient frameworks in networked system~\cite{nsdi24_grace,li2024reparolossresilientgenerativecodec} is that they assume the decoder can detect which indices are lost during transmission. This assumption holds in those frameworks because they packetize indices, making it possible to detect packet losses at the transport layer. In contrast, our approach transmits the entire set of indices in a single packet directly at the PHY layer, using signal processing techniques to conserve bandwidth. Hence, pinpointing corrupted indices becomes significantly challenging, as error detection capabilities at the PHY layer are inherently limited~\cite{hamming1986coding}.

\subsection{Packetization}
\label{sec:packetizer}

Recall from \S\ref{sec:mov_challenges} that underwater image transmission
can take tens of seconds due to limited bandwidth,
and is highly susceptible to transmission errors caused by channel variations at the PHY layer.
To address these challenges, we insert a training symbol into the image packet
after every $N$ data symbols. This allows for more frequent channel
equalization
to correct symbol distortion introduced by the channel, and facilitates
time synchronization to compensate for device movement.
Taken together, more training symbols help reduce the symbol error rate (SER).

\begin{figure}[t]
    \centering
    \begin{minipage}[h]{0.26\textwidth}
        \centering
    \includegraphics[width=\textwidth]{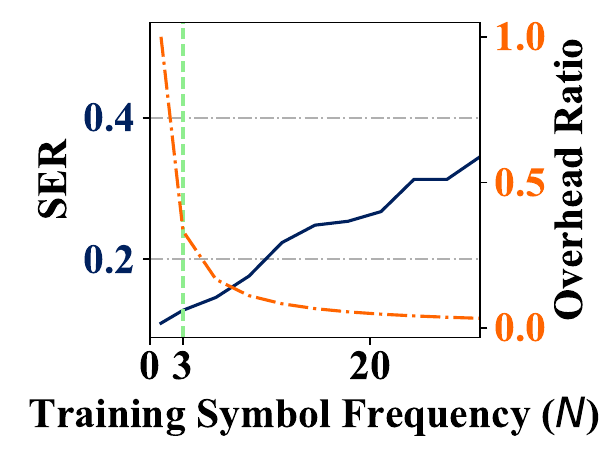}
    \vspace{-12pt}
    \caption{Symbol error rate (SER) and overhead ratio at
    varying training symbol frequencies.}
\label{fig:motiv_multi_training_symbols}
    \end{minipage}
    \hfill
    \begin{minipage}[h]{0.2\textwidth}
        \centering
    \includegraphics[width=\textwidth]{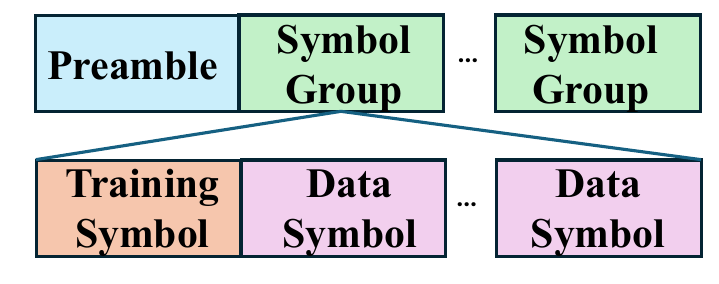}
   \vspace{-8pt}
    \caption{Image packetization.}
    \label{fig:image_packet}
    \end{minipage}
\end{figure}

However, as shown in Fig.~\ref{fig:motiv_multi_training_symbols},
increasing the number of training symbols (\ie reducing $N$)
also raises the overhead ratio, leading to longer transmission
and signal processing times.
To balance this trade-off, we set $N=3$, which ensures that
the duration of each symbol group\footnote{Given a symbol time
$T_{s}$ of 16 ms (as explained in \S\ref{sec:implementation}), the duration
of each symbol group is $(N+1) \times T_{s}  = 64$ ms.}
remains within the channel coherence time underwater~\cite{yang2012properties}.

In traditional RF-based wireless systems, multiple training symbols are commonly used in OFDM~\cite{3GPP}. However, as discussed in \S\ref{sec:mov_challenges},
the OFDM protocol encounters significant challenges in underwater image transmission,
primarily due to substantial SNR drops in initially selected subcarriers
during long mobile transmissions.
Consequently, adding multiple training symbols for equalization becomes ineffective,
as equalization fails when the SNR is too low.
Instead, we adopt the CSS (chirp spread spectrum) technique~\cite{xie2024icc,jia2022two,steinmetz2018practical,steinmetz2022taking,TSN23} to convert image packets into chirp signals
as it offers a balance between robustness and transmission speed.

Furthermore, we introduce a preamble to assist the receiver in detecting
the packet and determining its starting timestamp.
Reliable preamble detection at low SNR is critical in underwater environments;
otherwise, all other efforts become futile if the packet cannot be detected.
We adopt the preamble from \cite{chen2022underwater}, which exhibits strong
auto-correlation properties for robust detection.
Additionally, the detection algorithm has a lower false positive rate,
meaning it is less likely to be triggered by ambient underwater noise.

In summary,
each image packet consists of a single preamble, followed by multiple symbol groups,
where each symbol group includes one training symbol and three data symbols as shown in Fig.~\ref{fig:image_packet}.

\subsection{Smoothed \& Bounded Time Synchronization}\label{sec:synchronization}
The receiver's microphone continuously records audio and monitors for packet arrival
using the detection algorithm in~\cite{chen2022underwater}.
To identify packets, we perform cross-correlation between the received signal and
the preamble. Once a correlation peak is found,
an image packet is detected, and the receiver
captures sound for a fixed duration (slightly longer than the packet length)
as raw received data.
Nevertheless, due to the movement of mobile devices, symbol time offset is inevitable. %
To correct this offset, the receiver identifies the start timestamp of each training symbol via cross-correlation and adjusts the timestamps of the following data symbols.
However, directly applying cross-correlation for symbol timing synchronization is error-prone due to channel fading, multipath effects, and
ambient noise.

Therefore, to enhance the robustness of time synchronization,
we leverage two insights related to physical principles.
First, the relative speed between the sender and receiver during a specific
activity (e.g., scuba diving) is naturally constrained, \ie the difference in start timestamps between two consecutive training symbols is bounded:
\begin{align}
    | t_{tsym}^{i-1} + N_{sgroup}- t_{tsym}^{i} |  \leq \Delta,
\end{align}
where $t_{tsym}^{i}$ denotes the start index of the $i$-th training symbol,
and $N_{sgroup}$ represents the length of the sending symbol group in samples.
In our setup, $N_{sgroup} = (N+1) \times N_{s} = 3072$, where $N_{s}$ is the number of samples per symbol.
We set $\Delta = 40$ empirically based on preliminary experiments,
taking into account the maximum relative movement speed between the sender
and receiver, as well as orientation changes.
This bounded constraint allows us to efficiently search for the start
time of the next training symbol within a limited range,
thereby improving synchronization accuracy and computational efficiency
on mobile devices.

\begin{figure}[t]
    \centering
    \includegraphics[width=\linewidth]{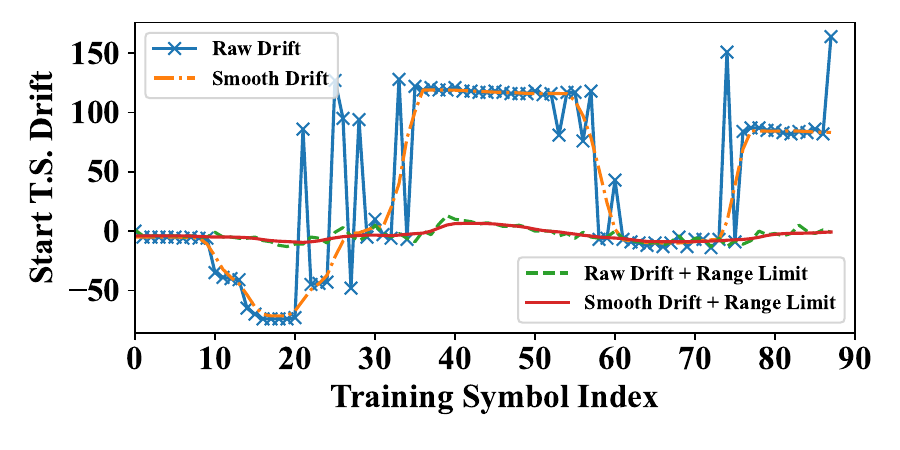}
    \vspace{-20pt} %
    \caption{Start timestamp drift in training symbols.}
    \label{fig:doppler_observation}
\end{figure}

Second, the distance between the sender and receiver should change gradually,
\ie the drift in the start timestamp of each training symbol should also
change smoothly. We calculate the drift as the difference between the
detected start timestamp of the training symbol and its expected start timestamp
($i \times N_{group}$) when there is no movement. If the drift changes
drastically, resulting in excessive jitter as shown by the blue line
in Fig.~\ref{fig:doppler_observation}, we can infer that
an error has occurred in symbol time synchronization.

In Fig.~\ref{fig:doppler_observation}, the x-axis represents the index of the training symbol, and the y-axis shows the start timestamp drift. We collected this data by submerging two phones underwater from two boats and driving the boats in random directions and at varying distances.
\textit{Raw Drift} does not address jitter specifically, leading to noisy and inaccurate predictions due to imperfections in the cross-correlation results.
Leveraging our second insight, we apply a moving average~\cite{moving_average} to smooth \textit{Raw Drift} and eliminate spikes, as shown by \textit{Smooth Drift}. While this method effectively corrects short-term drift (and thus identifies the correct start timestamp quickly), it fails to handle long-term drift (where the timestamps shift beyond the recoverable range). To address this, we further constrain drift within a limited range based on our first insight. This is denoted as \textit{Smooth Drift + Range Limit} and used to
recalculate the start timestamp of each training symbol. Later, we demonstrate that
our proposed drift smoothing and bounding approach significantly reduces
SERs.

After compensating for the influence of movement and smoothing out detection errors, we utilize the correct start timestamps of two consecutive training symbols---obtained from the previous symbol time synchronization---to segment the symbol group and extract the data symbols. Finally, we apply linear interpolation to the data symbols to mitigate the Doppler effect.

\section{Implementation}\label{sec:implementation}

\subsection{Generative Model}

\noindent\textbf{Dataset and model training.}
We curate two non-overlapping datasets for model training and testing.
The training dataset comprises 78,420 images sourced from public underwater
datasets~\cite{anantharajah2014local, marques2020l2uwe, islam2020suim,
islam2019fast, jmse7010016, 2018arXiv180704856G, li2019underwater, 10102831} and
135,000 general images sampled from ImageNet~\cite{imagenet100}.
For testing, we extract 1,216 underwater images from five scuba diving
videos~\cite{dataset_video_1, dataset_video_2, dataset_video_3, dataset_video_4,
dataset_video_5}, captured in various diving locations and conditions.

We initialize our model using a pre-trained TiTok model~\cite{yu2024image} and perform context-aware token distillation for 112 epochs on our training dataset,
followed by 60 epochs of error-resilient fine-tuning.
These two stages require 11 and 14 hours of training time on 4$\times$A100 GPUs,
respectively.

\parab{Model deployment.} Building upon the open-source code from~\cite{chen2022underwater}, we implement and deploy \sysname on the Android platform.
Our system follows the encoder-decoder architecture and training framework of TiTok,
where the codebook size is $K'=4096$, and a 256$\times$256 image is compressed into $M'=64$ tokens, equal to $64\times log_2{4096} = 768$ bits.
The neural network models, initially in PyTorch format, are converted into Android-compatible formats and optimized for mobile devices using PyTorch~\cite{pytorch_mobile} libraries.
With a total size of under 1 GB, these models can be loaded and executed
on modern smartphones, such as the Samsung S21~\cite{Samsung} used in our study.

\subsection{Transmitter}
The transmitter performs modulation and channel coding, converting serialized image bits from the image encoder into time-series acoustic signals. These signals are then broadcast into the water via the mobile device's speaker.
To ensure a high SNR for acoustic signals,
we select the 1.5--3.5 kHz frequency range for image transmission,
based on the measurement study presented in Fig.~\ref{fig:snr}.

\parab{Modulation.}
The CSS modulation technique encodes each symbol as a chirp signal that spans the entire bandwidth~\cite{TSN23,jung2021exploiting}. The chirp generation process is governed by two key parameters: the spreading factor ($SF$) and the bandwidth ($BW$).
To balance robustness and data rate, we set $SF$ $=$ 5,
achieving a data rate of 312 bps with $BW$ $=$ 2 kHz in underwater environments.
Consequently, the symbol time is $T_{s} = 2^{SF} / BW$ $=$ 16 ms.
After modulation, the mobile device transmits the acoustic signal at a sampling rate of
$f_{s} = 48$ kHz, generating $N_{s} = 768$ samples per symbol.

\parab{Channel coding.}
To enhance robustness and facilitate error recovery,
we employ encoding techniques commonly used in chirp-based communication systems~\cite {TSN23,LoRaWAN,chen2024hitting}:

\begin{itemize}[leftmargin=*]
    \item \textit{Hamming coding.} We first apply Hamming encoding~\cite{hamming1986coding} with a 4/7 code rate, encoding every 4 data bits with 3 parity check bits to form a 7-bit codeword.
    \item \textit{Diagonal interleaving.} We group every $SF$ $=$ 5 codewords into 7 symbols,
    distributing each bit of a codeword diagonally across 7 different
    symbols~\cite{TSN23}.
    This approach aligns with the fact that a (7, 4) Hamming code can correct up to 1-bit error per codeword.
    \item \textit{Gray coding.} %
    We use Gray coding~\cite{doran2007gray} to assign binary values to symbols so that adjacent
    symbols differ by only one bit. This minimizes the impact of demodulation errors
    and improves the effectiveness of Hamming coding in detecting and correcting bit errors.
\end{itemize}

\subsection{Receiver}
The receiver performs several key functions, including packet detection,
time synchronization, equalization, demodulation, and decoding,
to extract data bytes from the received raw signal.
These data bytes are subsequently forwarded to the image decoder for image
reconstruction.

\parab{Equalization.}
We apply MMSE-based time-domain equalization~\cite{MMSE} to each symbol group.
The equalization coefficients are estimated using each training symbol and then
applied to equalize the subsequent data symbols within the same group.
To balance performance and computational complexity,
we configure the equalizer with a tap size of 240 samples and an offset of 80 samples.
Furthermore, the time-domain equalization process helps mitigate symbol time
offsets during convolution operations.

\parab{Demodulation.}
After extracting each data symbol from the received packet, we apply a standard
de-chirp process to demodulate the symbol, following the approach in~\cite{TSN23}.

\parab{Channel decoding.}
The channel decoder applies the reverse process of the channel encoder,
incorporating Gray coding, diagonal de-interleaving, and Hamming decoding.
Once the data bytes are successfully decoded, they are passed to the image decoder
for final image reconstruction.

\section{Evaluation}
\label{sec:evaluation}

\begin{figure*}[t]
    \centering
    \includegraphics[width=\linewidth]{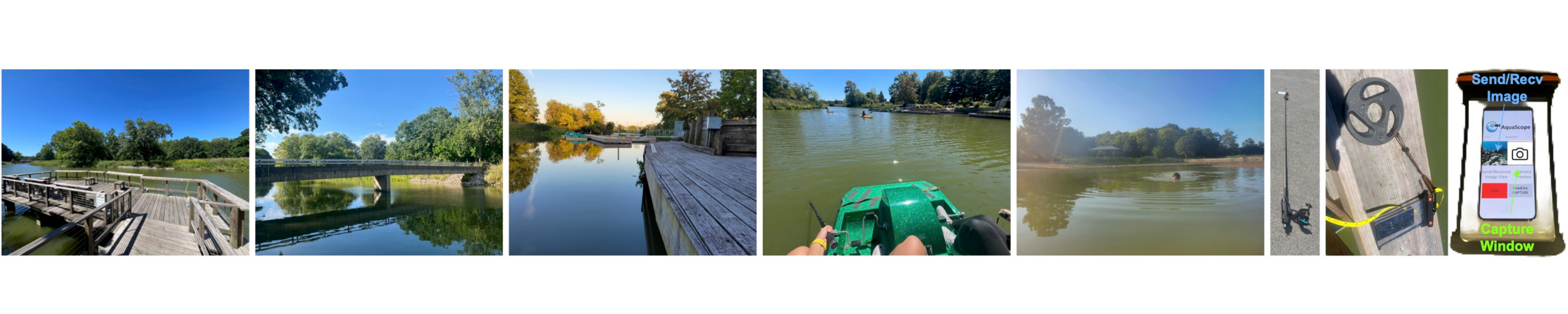}
    \vspace{-30pt}
    \caption{Experimental environments from left to right: fishing dock,
    bridge, lake bank, pedal boat, and lake beach. The remaining images show
    a prototype of \sysname and the mounting setup used to secure phones
    to the measurement tool when close to and when far from the water.}
\label{fig:experiment_environment}

\end{figure*}

\subsection{Experimental Setup}
To thoroughly evaluate our system, we conducted a series of tests in various environments using customized setups.

\parab{Experimental environments.}
We tested our system across different environments to assess the impact of distance, depth, orientation, mobility, and any other environmental factors on its performance. The environments considered are listed below and the corresponding images are shown in Fig.~\ref{fig:experiment_environment}.

\begin{itemize}[leftmargin=*]
    \item \em{Fishing dock:} A quiet environment with limited space for setup, suitable for short-range experiments.
    \item \em{Bridge:} A realistic and quiet environment with ample space
    (a measurement distance of up to 20 m and an average water depth of 3 m),
    making it the primary test site.
    \item \em{Lake bank:} A quiet site that allows for easy control of the phone's orientation; bank side has a reflective surface.
    \item \em{Pedal boat:} A busy and noisy location with kayaks and pedal boats passing by; may simulate realistic underwater movement patterns.
    \item \em{Lake beach:} A quiet site for testing real underwater image transmission between two swimmers.
\end{itemize}

As shown in Fig.~\ref{fig:experiment_environment},
when operating far from water, we secure the phone pouch to a tape measure hook with a climbing lock. A barbell weight, attached to the same hook with a  0.5m rope, ensures the phone stays submerged. The tape measure works as both a rope and a depth gauge. The phone in the pouch floats vertically in the water, maintaining an upright orientation when it is static.
When operating close to water, we attach the phone in its pouch to a three-meter-long selfie stick using a 0.5m rope, causing the phone's depth and orientation to vary randomly during the test.

The choice of the waterproof tool is essential for transmission quality and phone safety. After evaluating several options, as shown in Fig.~\ref{fig:experiment_environment}, we choose a pouch for its minimal signal attenuation and reliable waterproofing (rated to withstand 30 hours at a depth of 100 feet)~\cite{TORRAS}.

\captionsetup[subfigure]{aboveskip=0pt, belowskip=1pt} %
\begin{figure}[t]
    \centering
    \begin{minipage}{0.06\textwidth}
        \centering
        \raisebox{12\height}{\scriptsize \textbf{Image \#1}}
    \end{minipage}%
    \begin{subfigure}[b]{0.075\textwidth}
        \includegraphics[width=\linewidth]{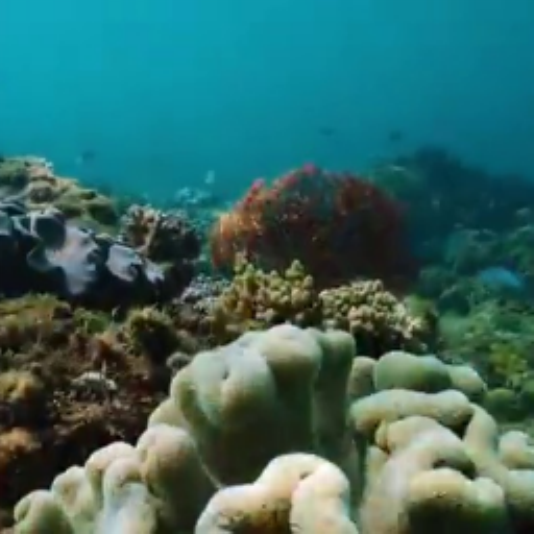}
        \caption*{\scriptsize Original}
    \end{subfigure}
    \begin{subfigure}[b]{0.075\textwidth}
        \includegraphics[width=\linewidth]{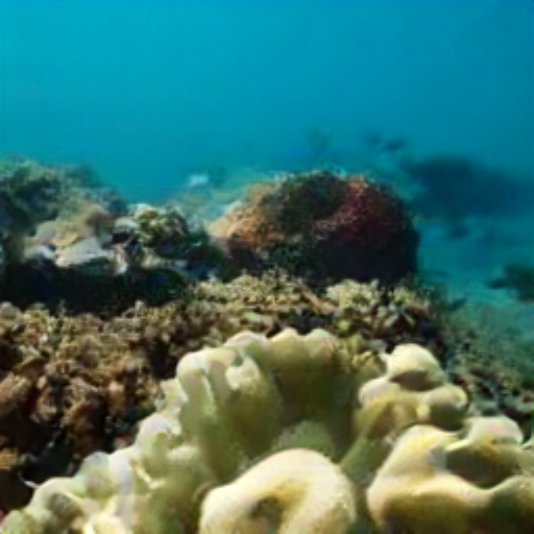}
        \caption*{\scriptsize 1.5\% 0.221}
    \end{subfigure}
    \begin{subfigure}[b]{0.075\textwidth}
        \includegraphics[width=\linewidth]{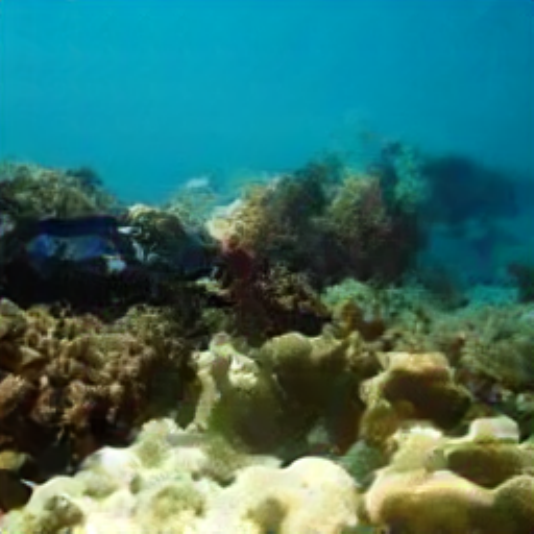}
        \caption*{\scriptsize 14\% 0.275}
    \end{subfigure}
        \begin{subfigure}[b]{0.075\textwidth}
        \includegraphics[width=\linewidth]{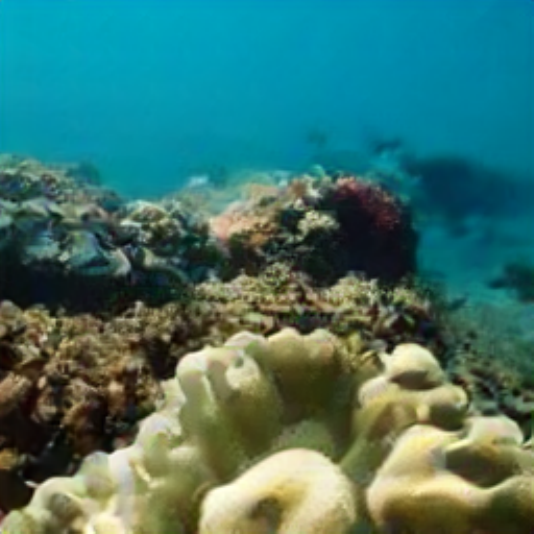}
        \caption*{\scriptsize 1.5\% 0.238}
    \end{subfigure}
        \begin{subfigure}[b]{0.075\textwidth}
        \includegraphics[width=\linewidth]{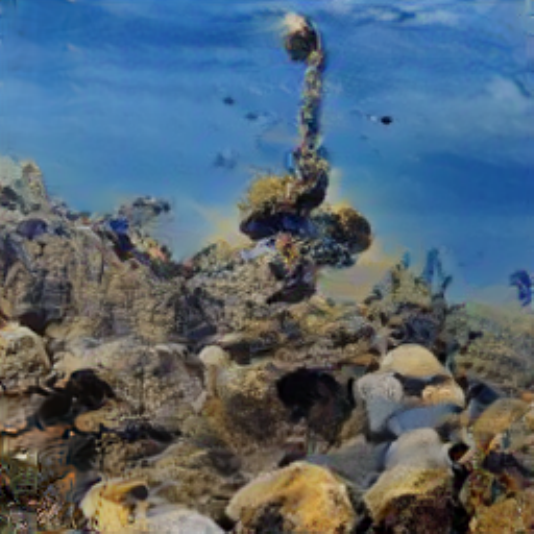}
        \caption*{\scriptsize 55\% 0.507}
    \end{subfigure}

    \vspace{-2.7em}
    \begin{minipage}{0.06\textwidth}
        \centering
        \raisebox{12\height}{\scriptsize \textbf{Image \#2}}
    \end{minipage}%
    \begin{subfigure}[b]{0.075\textwidth}
        \includegraphics[width=\linewidth]{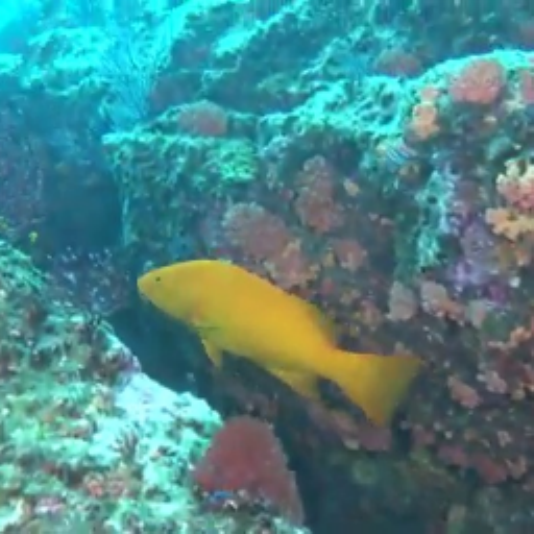}
        \caption*{\scriptsize Original}
    \end{subfigure}
    \begin{subfigure}[b]{0.075\textwidth}
        \includegraphics[width=\linewidth]{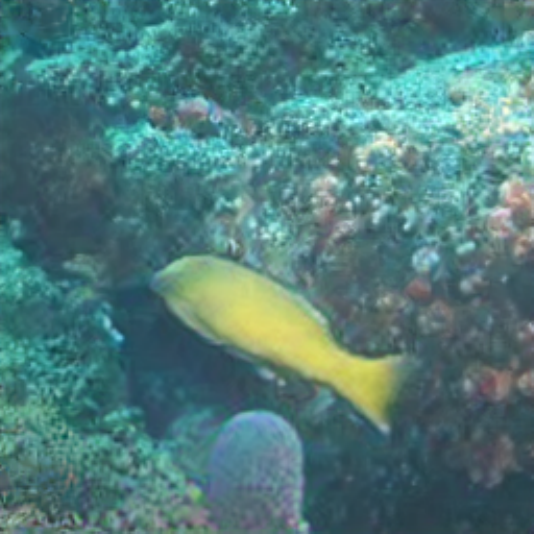}
        \caption*{\scriptsize 19\% 0.327}
    \end{subfigure}
    \begin{subfigure}[b]{0.075\textwidth}
        \includegraphics[width=\linewidth]{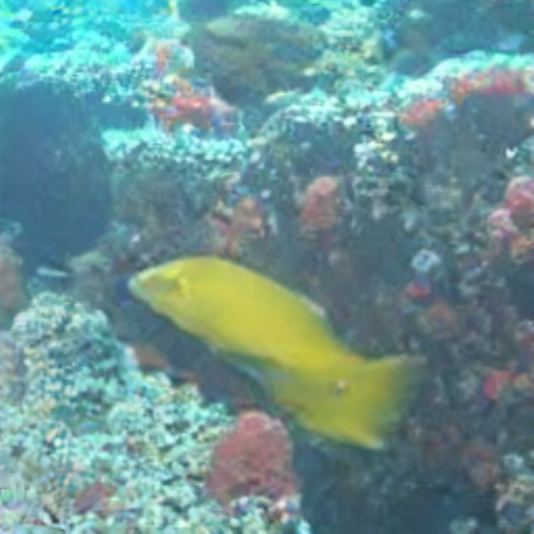}
        \caption*{\scriptsize 9\% 0.321}
    \end{subfigure}
        \begin{subfigure}[b]{0.075\textwidth}
        \includegraphics[width=\linewidth]{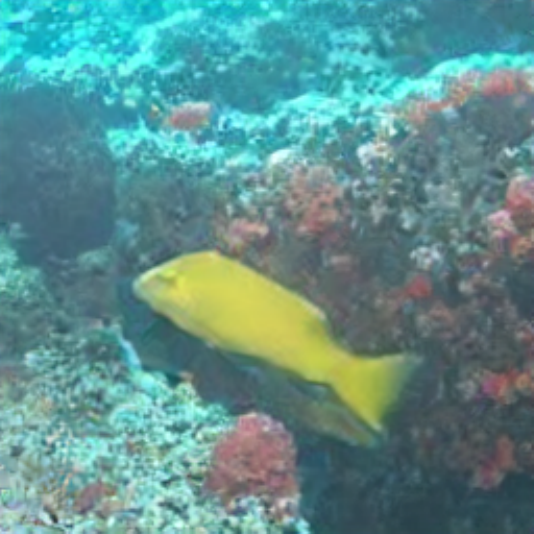}
        \caption*{\scriptsize 1.6\% 0.273}
    \end{subfigure}
        \begin{subfigure}[b]{0.075\textwidth}
        \includegraphics[width=\linewidth]{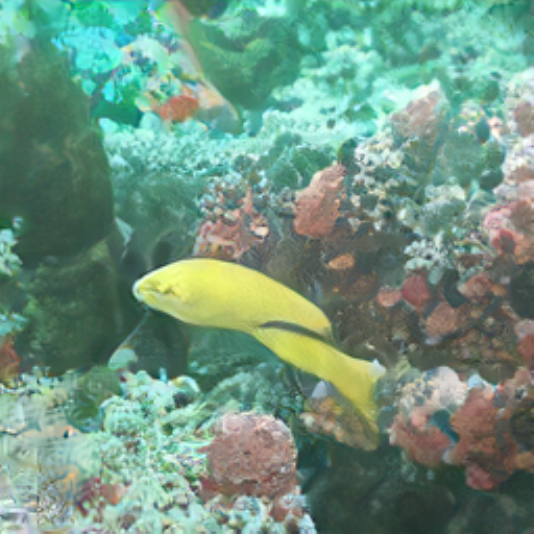}
        \caption*{\scriptsize 20\% 0.412}
    \end{subfigure}

    \vspace{-2.7em}
    \begin{minipage}{0.06\textwidth}
        \centering
        \raisebox{12\height}{\scriptsize \textbf{Image \#3}}
    \end{minipage}%
    \begin{subfigure}[b]{0.075\textwidth}
        \includegraphics[width=\linewidth]{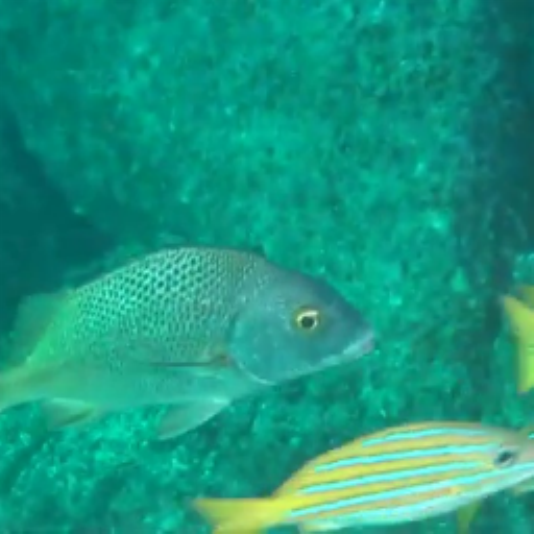}
        \caption*{\scriptsize Original}
    \end{subfigure}
    \begin{subfigure}[b]{0.075\textwidth}
        \includegraphics[width=\linewidth]{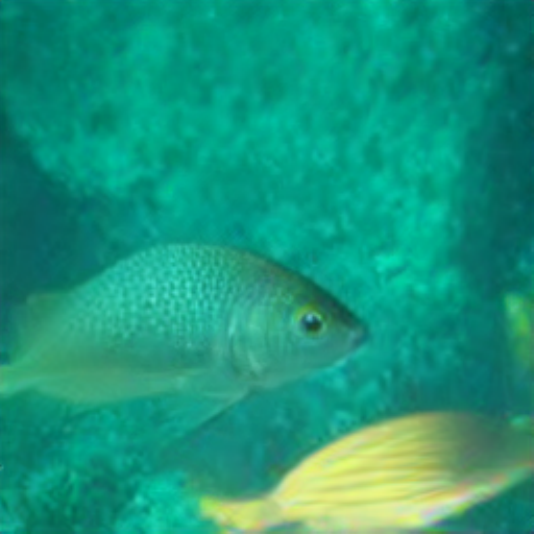}
        \caption*{\scriptsize 3\% 0.183}
    \end{subfigure}
    \begin{subfigure}[b]{0.075\textwidth}
        \includegraphics[width=\linewidth]{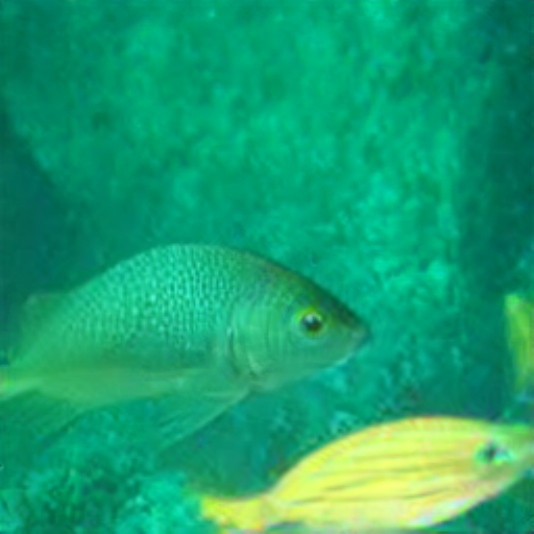}
        \caption*{\scriptsize 3\% 0.189}
    \end{subfigure}
        \begin{subfigure}[b]{0.075\textwidth}
        \includegraphics[width=\linewidth]{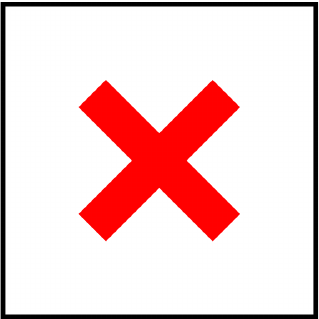}
        \caption*{\scriptsize MISS}
    \end{subfigure}
        \begin{subfigure}[b]{0.075\textwidth}
        \includegraphics[width=\linewidth]{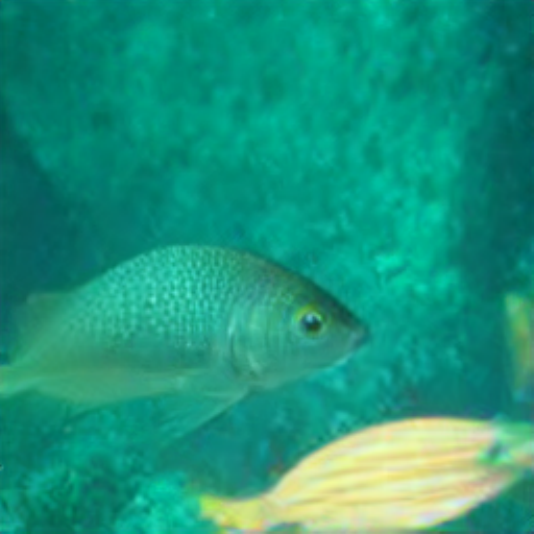}
        \caption*{\scriptsize 0\% 0.177}
    \end{subfigure}

    \vspace{-2.7em}
    \begin{minipage}{0.06\textwidth}
        \centering
        \raisebox{12\height}{\scriptsize \textbf{Image \#4}}
    \end{minipage}%
    \begin{subfigure}[b]{0.075\textwidth}
        \includegraphics[width=\linewidth]{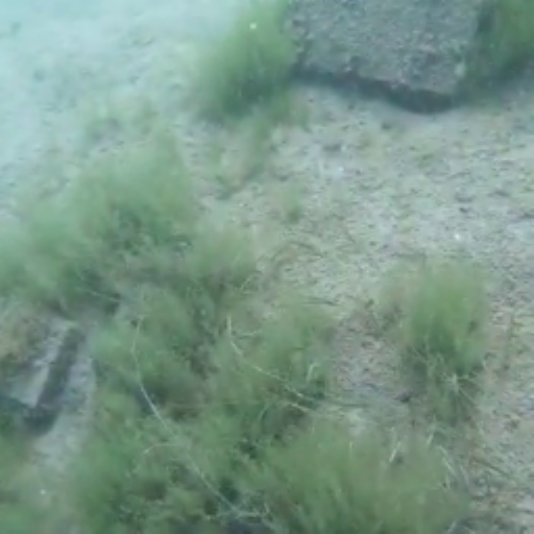}
        \caption*{\scriptsize Original}
    \end{subfigure}
    \begin{subfigure}[b]{0.075\textwidth}
        \includegraphics[width=\linewidth]{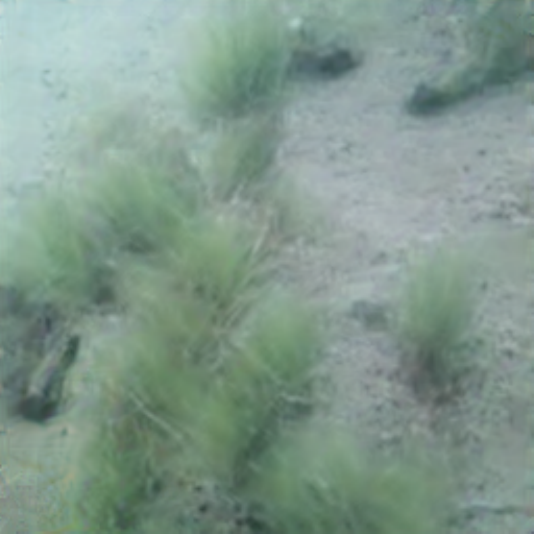}
        \caption*{\scriptsize 6\% 0.340}
    \end{subfigure}
    \begin{subfigure}[b]{0.075\textwidth}
        \includegraphics[width=\linewidth]{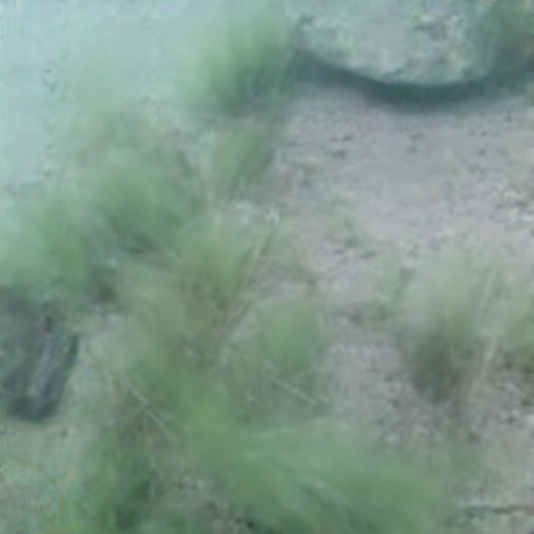}
        \caption*{\scriptsize 11\% 0.339}
    \end{subfigure}
        \begin{subfigure}[b]{0.075\textwidth}
        \includegraphics[width=\linewidth]{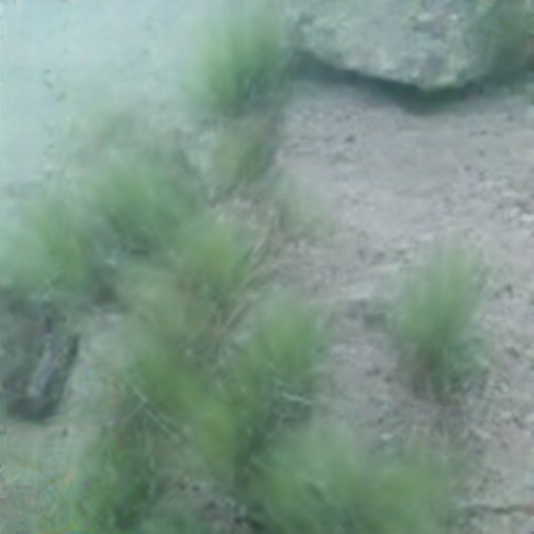}
        \caption*{\scriptsize 0\% 0.320}
    \end{subfigure}
        \begin{subfigure}[b]{0.075\textwidth}
        \includegraphics[width=\linewidth]{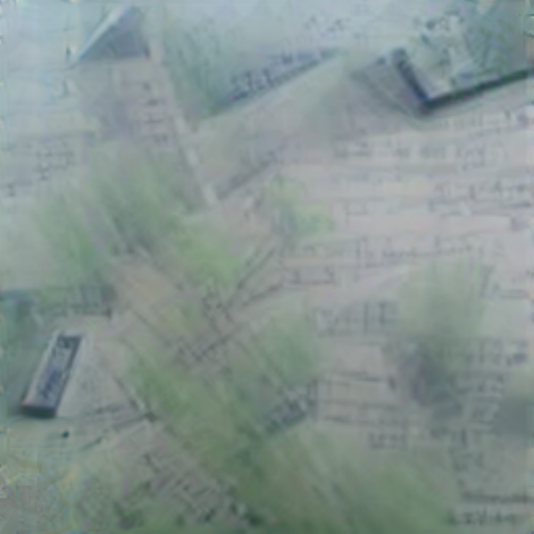}
        \caption*{\scriptsize 14\% 0.502}
    \end{subfigure}
    \vspace{-2.7em}
    \caption{End-to-end performance of \sysname. For each reconstructed image, the caption displays the IER (percentage, average 10.5\%) and LPIPS (second value, average 0.30).}
    \label{fig:eval_overall}

\end{figure}

\parab{Baseline and metrics.}
We consider AquaApp~\cite{chen2022underwater} as one of our baseline systems. AquaApp is originally designed for text transmission using OFDM protocol. We mainly compare our system with the transmission protocol part of AquaApp. We also implement a conventional CSS-based acoustic system based on~\cite{TSN23} as another baseline system. Specifically, it leverages additional 4 symbol chirps (2 up-chirps and 2 down-chirps) for symbol time synchronization.
We integrate both baseline systems with the generative image codec to make them comparable to \sysname. We refer them to as ``OFDM'' and ``CSS'' for simplicity in the following discussions.

We consider two metrics when evaluating \sysname and baseline schemes (unless otherwise stated).
First, we use \textit{Index Error Rate (IER)} as one performance metric since the image quality is directly determined by the correctness of indices, and it also reflects the effectiveness of the techniques at the PHY layer.
Second, we use \textit{LPIPS} as another metric to evaluate the fidelity of the received image. The image metric should reflect if the receiver can understand the semantic meaning of the original image from the received image.
To find such a metric, we manually score the received images with reference to the original image, categorizing them as cannot understand, moderate understanding, and well understanding. We then attempt to find correlations with the following image metrics: PSNR, SSIM~\cite{hore2010image}, MS-SSIM~\cite{wang2003multiscale}, VIF~\cite{sheikh2005visual}, LPIPS, CLIP score~\cite{radford2021learning}, and DinoV2 score~\cite{oquab2023dinov2}. CLIP and DinoV2 scores are cosine similarities between embeddings of received and ground truth images, calculated by large vision models known for semantic representation extraction. Among all the metrics, LPIPS shows the highest correlation, at $-0.4$. Hence, we use LPIPS as our image metric.

\begin{figure*}[t]
    \centering
    \includegraphics[width=\linewidth]{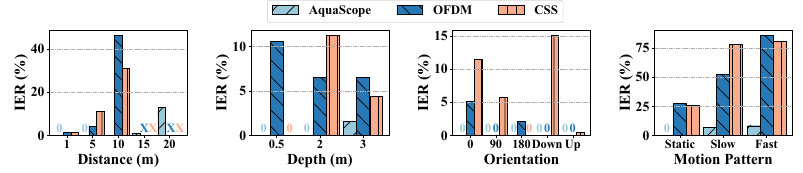}
    \caption{Impact of different physical factors. An ``X'' mark indicates that the receiver does not receive the image packets.}
    \label{fig:eval_different_factors}
\end{figure*}

\subsection{Evaluation Results in the Wild}\label{sec:eval_results}

\noindent\textbf{End-to-end performance.}
Fig.~\ref{fig:eval_overall} shows the end-to-end performance of \sysname in the real underwater scenario. We perform the experiment in a lake. During the experiment, two people swim and tread water at random directions and speeds with mobile phones held in hand to mimic the real scuba diving scenario.
In the test, the sender sends four diverse images (shown in each row), and each image is sent 4 times (received images shown in each column).  15 out of 16 sent images are successfully received. The missing one is due to the failure of preamble detection. 14 out of the 15 received images (except for the last image in the first row) maintain their original meaning, proving \sysname's capability in reliably delivering high-fidelity images in realistic diving scenarios. Additionally, the decoder shows great recovery ability. For example, comparing the second and third images in the first row, we can find that although IER jumps from $1.5\%$ to $14\%$, the recovered image still maintains most of the semantic meaning. The success of transmitting diverse images also proves that \sysname's generalizability to different underwater scenarios such as transmitting simple (Image \#2) or complex (Image \#1) objects.

\parab{Effect of distance.} The first image in Fig.~\ref{fig:eval_different_factors} shows IER versus distance. We have several observations. When the distance between sender and receiver is small ($<$5m), all schemes have a small IER which is below 11\% (2\% BER). In particular, \sysname has no index error.
When the distance is between 10--15m, \sysname still shows great performance of low IER while IERs of CSS and OFDM surges to 31\% (6\% BER) and 46\% (19\% BER), respectively.
Even at 20m, the IER of \sysname is tolerable. The mark ``X'' in the figure indicates that the image packets are not received at the receiver side. Based on the result, \sysname can reliably deliver images with a distance of up to 20m, and the performance starts to decrease when the distance is larger than 20m due to signal attenuation.

\parab{Effect of depth and orientation.}
The second and third images in Fig.~\ref{fig:eval_different_factors} show IER versus depth and orientation, respectively. For depth, we placed the mobile phones close to the water surface, at the middle (2m) and the bottom (3m) of the lake, respectively, with a distance of 5m.
For orientation, we first set the speaker of the sender and the microphone of the receiver to face each other. Then, we rotate the receiver phone in the azimuth angle from $0^{\circ}$ to $90^{\circ}$ and $180^{\circ}$, respectively. Furthermore, we also rotate the pitch angle of the receiver phone to $90^{\circ}$ and $-90^{\circ}$ (named up and down) to test the effect when the speaker and microphone are not in the same plane. We mount the phone to a phone holder connected to a selfie stick and submerge the phone underwater to about 1.5m depth with our hands holding the stick.
The results are shown in Fig.~\ref{fig:eval_different_factors}, we can see that depth and orientation changes of the mobile phones have less impact on the systems and IER is generally below 15\% (2\% BER). The variance in the result may come from the slight movement during the experiment.

\parab{Effect of mobility.}
The last image in Fig.~\ref{fig:eval_different_factors} shows IER versus different mobility at the distance of 5m. We submerge two mobile phones connected to tape measures from the bridge in the lake and keep the sender still. We randomly move the receiver by moving one end of the tape measure moderately and intensively.
As depicted in the figure, \sysname shows great resilience to the mobility and maintains IER within 8\% (1.5\% BER) while over 50\% indices are wrong in CSS and OFDM when the mobile devices move.

\subsection{Deep Dive}

\begin{figure}[t]
    \centering
    \includegraphics[width=0.5\textwidth]{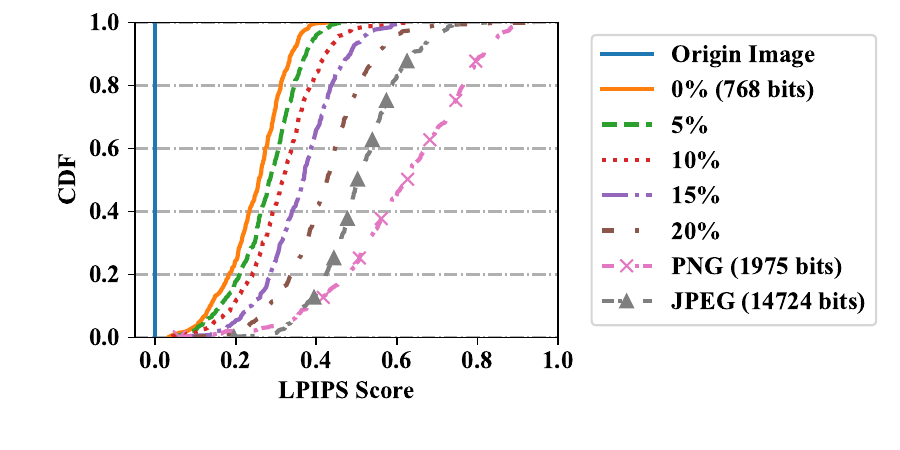}
    \vspace{-8mm}
    \caption{LPIPS achieved by \sysname at different IERs \emph{vs.}
    traditional codecs (PNG and JPEG).}
    \label{fig:ByERvsLPIPS}
\end{figure}

\noindent\textbf{Codecs comparisons.}
We compare traditional image codecs with a generative codec and analyze the impact of token errors. The results from experiments conducted on 400 high-quality images from the test dataset are shown in Fig.~\ref{fig:ByERvsLPIPS}. Our observations indicate that as the IER increases, the LPIPS score also rises, signifying a loss of visual details in the received image. Notably, even in the absence of transmission errors and despite having a much larger payload size, the two traditional image codecs perform significantly worse than \sysname.

\begin{figure*}[t]
    \centering
    \begin{subfigure}[b]{.24\linewidth}
        \centering
        \includegraphics[width=\linewidth]{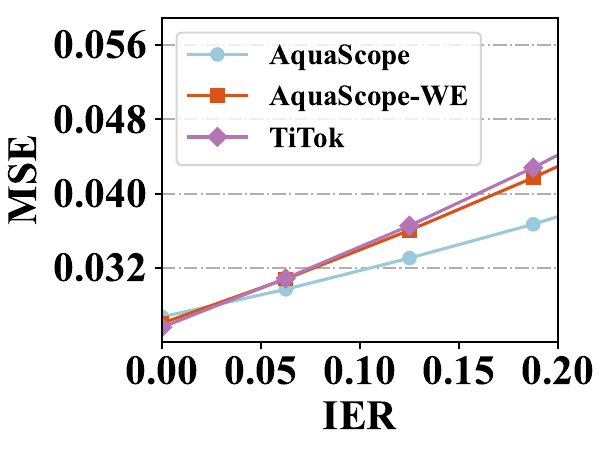}
        \label{fig:MSE_VS_IER_fine_tune}
    \end{subfigure}
    \begin{subfigure}[b]{.24\linewidth}
        \centering
        \includegraphics[width=\linewidth]{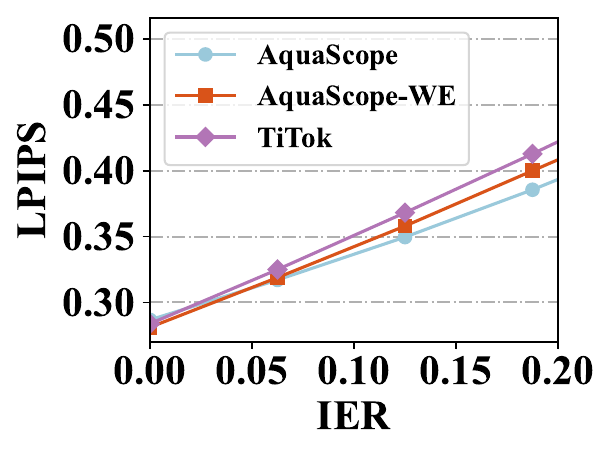}
        \label{fig:LPIPS_VS_IER_fine_tune}
    \end{subfigure}
        \begin{subfigure}[b]{.24\linewidth}
        \centering
        \includegraphics[width=\linewidth]{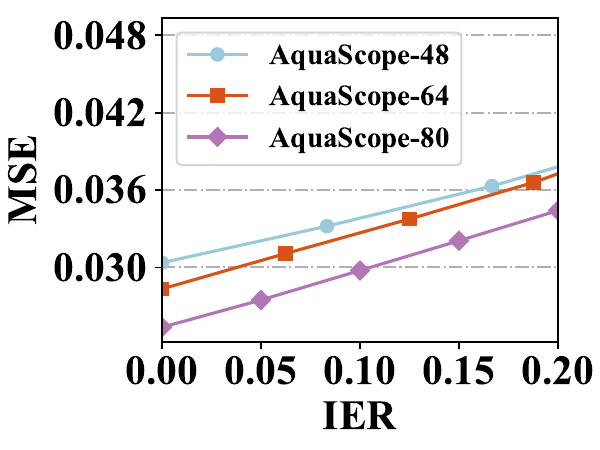}
        \label{fig:MSE_VS_IER_token}
    \end{subfigure}
    \begin{subfigure}[b]{.24\linewidth}
        \centering
        \includegraphics[width=\linewidth]{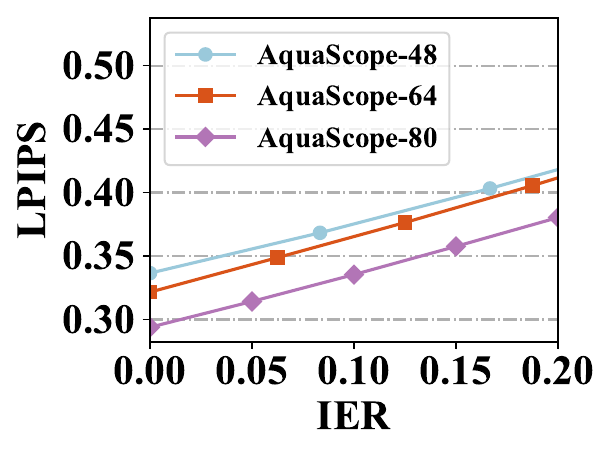}
        \label{fig:LPIPS_VS_IER_token}
    \end{subfigure}
    \vspace{-5mm}
    \caption{Impact of fine-tuning and the number of distilled tokens $M'$
    on reconstructed image quality.}
    \label{fig:eval_fine_tuning}
    \vspace{3pt}
\end{figure*}

\parab{Fine-tuning.}
To illustrate the advantage of error-resilient fine-tuning, we consider two new baselines: TiTok~\cite{yu2024image} and \sysname-WE. TiTok is the state-of-the-art generative model trained on ImageNet~\cite{imagenet15russakovsky} and \sysname-WE is fine-tuned on our collected underwater image dataset without introducing transmission errors. We simulate the IER from 0\% to 20\% and the results are shown in Fig.~\ref{fig:eval_fine_tuning}. We find that as the IER increases, the LPIPS score and mean square error (MSE) of all approaches also rise. Furthermore, \sysname can achieve up to 8\% and 21\% gains in the LPIPS and MSE, respectively, compared to the baselines. This demonstrates \sysname is more robust to the transmission errors after fine-tuning.

\parab{Distilled token numbers.}
We train different transformer versions with 48, 64, and 80 distilled tokens from scratch, namely \sysname-48, \sysname-64, and \sysname-80, respectively.
The results are shown in Fig.~\ref{fig:eval_fine_tuning}. \sysname-80 has better LPIPS and MSE as it compresses the image with more (redundant) tokens which is more robust to the indice errors. To balance robustness and computation and communication overhead, we choose $M'$ = 64 distilled tokens.

\begin{figure}[t]
    \centering
    \begin{subfigure}[b]{0.09\textwidth}
        \includegraphics[width=\linewidth]{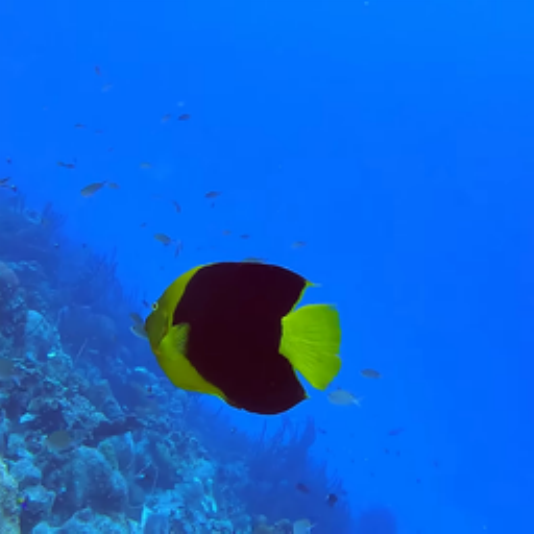}
        \caption*{\scriptsize Original}
    \end{subfigure}
    \begin{subfigure}[b]{0.09\textwidth}
        \includegraphics[width=\linewidth]{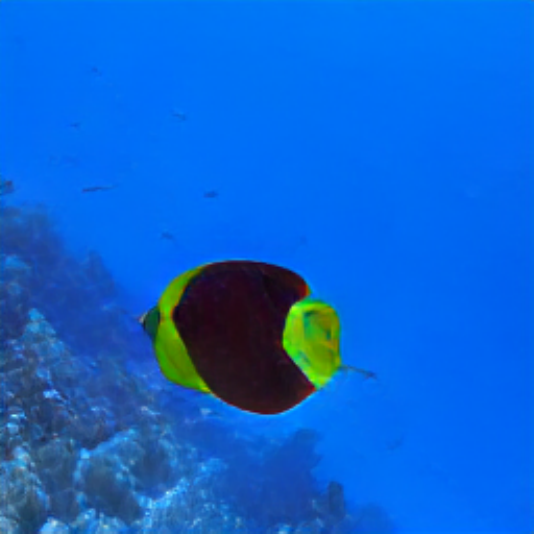}
        \caption*{\scriptsize VQGAN (256)}
    \end{subfigure}
    \begin{subfigure}[b]{0.09\textwidth}
        \includegraphics[width=\linewidth]{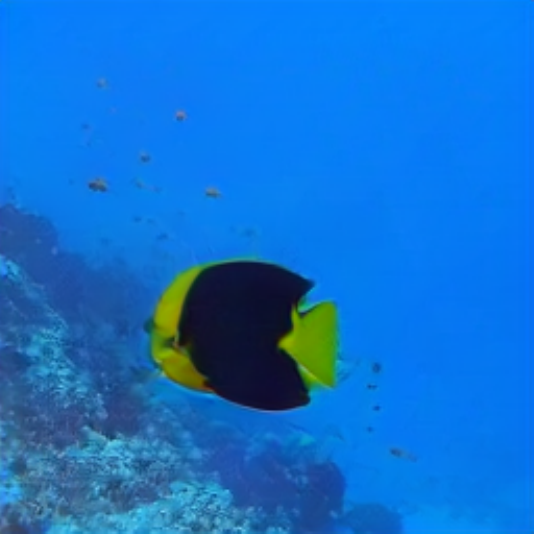}
        \caption*{\scriptsize Distilled (64)}
    \end{subfigure}
        \begin{subfigure}[b]{0.09\textwidth}
        \includegraphics[width=\linewidth]{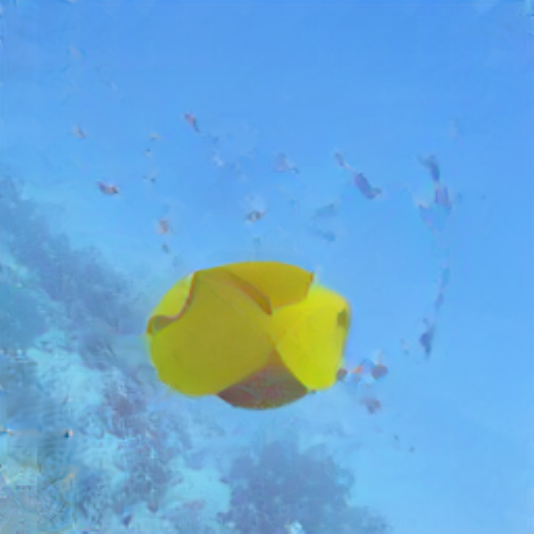}
        \caption*{\scriptsize TiTok (20\%)}
    \end{subfigure}
    \begin{subfigure}[b]{0.09\textwidth}
        \includegraphics[width=\linewidth]{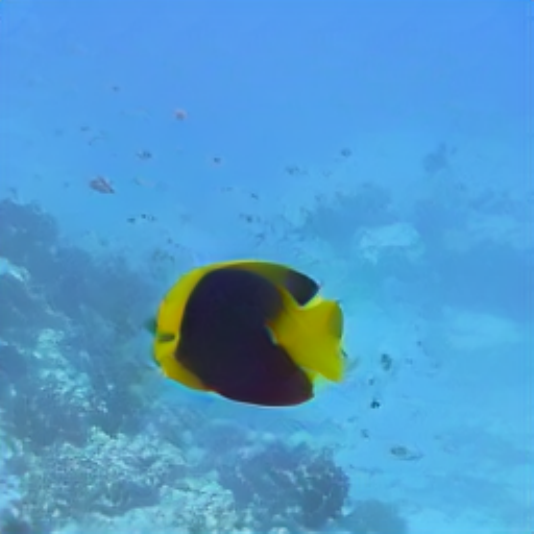}
        \caption*{\scriptsize Ours (20\%)}
    \end{subfigure}
    \caption{Samples of reconstructed images from different methods. VQGAN uses 256 tokens. ``Distilled'' is our fine-tuned version of TiTok without transmission errors. TiTok employs the base-64 model with a 20\% IER. ``Ours'' represents our fine-tuned version of TiTok with a 20\% IER. %
    }
    \label{fig:semantic_compression_filtering_intuition}
    \vspace{3pt}

\end{figure}

\parab{Visualization examples.} We compare the reconstructed images from different methods in Fig.~\ref{fig:semantic_compression_filtering_intuition}.
Our first observation is that the image reconstructed from distilled tokens exhibits similar or even better visual quality than the one reconstructed using the VQGAN model, highlighting the importance of context-aware customization for the generative models.
Our second observation is that when the perturbed tokens are identical—where we control both the perturbed tokens and their values to be the same in the last two images—our model, leveraging the recovery capability of error-resilient fine-tuning, retains many visual details. In contrast, the baseline images lose both the visual details and the semantic meaning of the original image.

\captionsetup[subfigure]{aboveskip=0pt, belowskip=1pt} %

\parab{Time equalization and synchronization.}
We examine the importance of reliability enhancement techniques at the PHY layer proposed by \sysname, as illustrated in Fig.~\ref{fig:eval_ablation}. The SER is used as the evaluation metric because it accurately reflects errors (before channel decoding) in the chirp-based communication system. When only a single equalization is applied at the beginning of the entire image packet (referred to as ``one Equal.''), the system exhibits an average SER exceeding 50\%, indicating that a single initial equalization is insufficient to account for the evolving channel conditions in packets lasting several seconds. Introducing multiple training symbols for repeated equalization, but without employing symbol time synchronization (referred to as ``wo Sync.''), reduces the average SER to below 20\%. Utilizing cross-correlation to estimate the start time of each training symbol, but without applying the smoothed and bounded time offset correction (referred to as ``wo S+B''), mitigates the time offsets caused by movement, further lowering the SER to 11\%. Finally, applying the smoothed and bounded time offset correction, as described in \S\ref{sec:synchronization}, reduces the SER by an additional 2\%, sometimes correcting up to five additional index errors.

\subsection{System-level Performance}\label{sec:eval_system_metrics}

\begin{figure*}[t]
    \centering
    \begin{minipage}{0.33\textwidth}
        \centering
        \includegraphics[width=\textwidth]{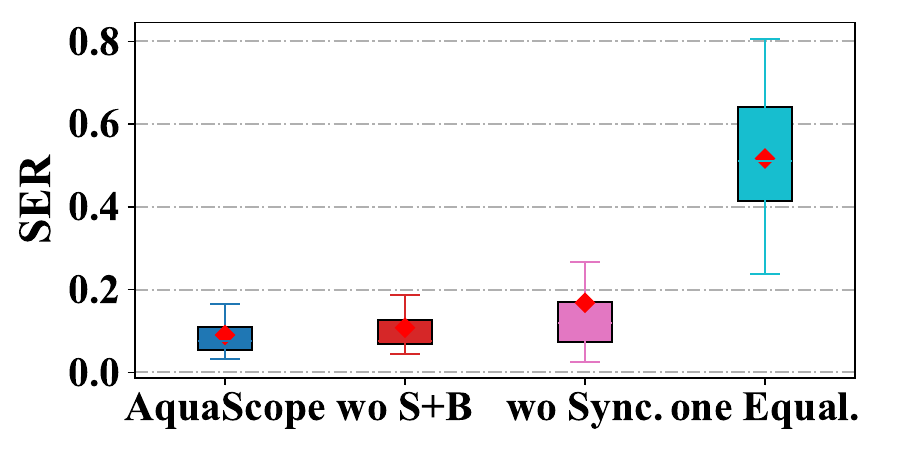}
        \caption{Ablation results at PHY.}
        \label{fig:eval_ablation}
    \end{minipage}\hfill
    \begin{minipage}{0.33\textwidth}
        \centering
        \includegraphics[width=\textwidth]{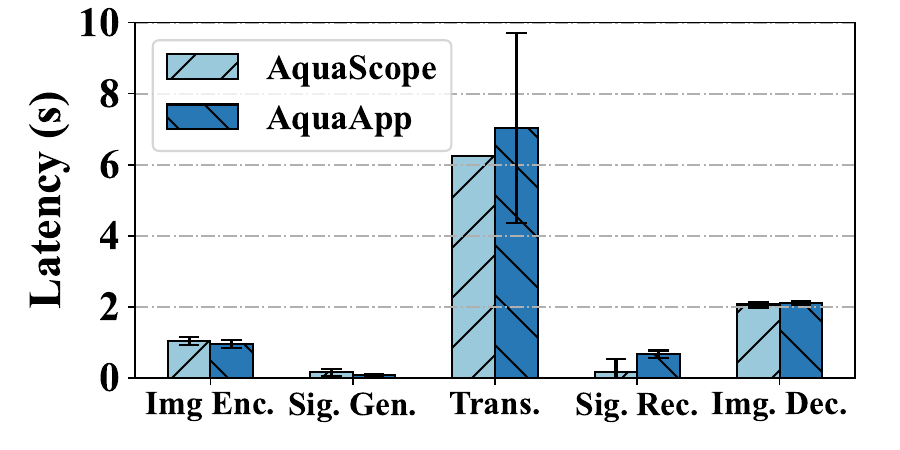}
        \caption{Latency breakdown.}
        \label{fig:latency_breakdown}
    \end{minipage}\hfill
    \begin{minipage}{0.33\textwidth}
        \centering
        \includegraphics[width=\textwidth]{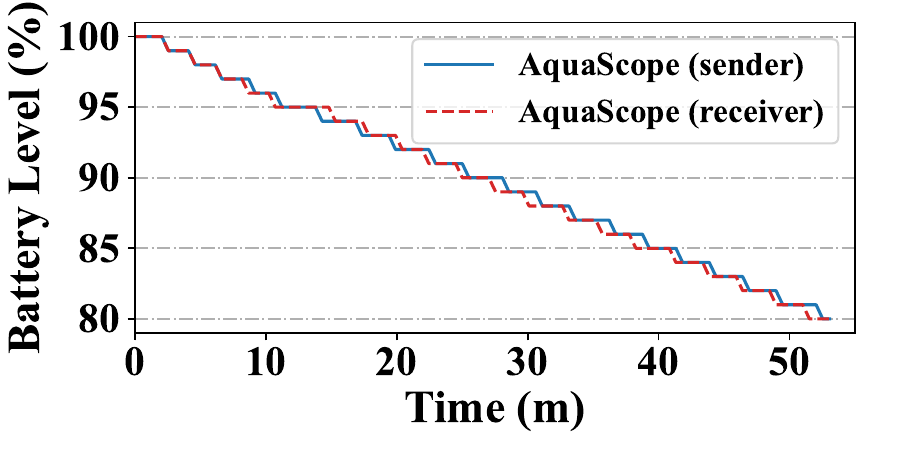}
        \caption{Battery drain over time.}
        \label{fig:battery}
    \end{minipage}
\end{figure*}

\noindent\textbf{Latency breakdown.}
The averaged end-to-end latency of \sysname is 9.2s with a standard deviation (std) of 0.66s. The average latency of the AquaApp is 10.9s and the std is 3.0s. Fig.~\ref{fig:latency_breakdown} shows latency breakdown of \sysname and AquaApp that contains image encoding, signal generation including channel coding and modulation, transmission, signal recovery including demodulation and channel decoding, and image decoding latency. We found transmission latency is the bottleneck of the end-to-end system latency. Compared to \sysname, AquaApp has much more variations in transmission latency as its data rate changes according to varying channel quality.

\parab{Energy.}
Fig.~\ref{fig:battery} illustrates the battery percentage drained at both the sender and receiver over approximately $50$ minutes of continuous image transmission. The results indicate that the battery level decreases by only 20\%, suggesting that the device could last for over three hours on a full charge. In contrast, recreational diving, typically limited by the size of the gas tank, lasts for up to an hour~\cite{mantacabo_scubadiving}, which is significantly shorter than the battery life of a mobile phone when using \sysname. Hence, \sysname is practical for recreational diving and can also support professional diving which lasts for a longer time.

\section{Discussion}
\label{sec:discussion}
In this section, we discuss the limitations of \sysname and opportunities for improvement.

\parab{Non-line-of-sight.} Currently, our system experiences performance drops when the sender and receiver are in a non-line-of-sight condition, primarily due to severe multipath effects and attenuation.
The obstruction can be caused by the human body (whether it involves the sender, receiver, or both) or by large obstacles like coral or shipwrecks that completely block the line of sight between the sender and receiver.
For the first type of obstruction, %
one possible solution is that
the sender and receiver (upon hearing the transmission sound) can opt to expose the mobile phone to reduce the obstruction. The second type of obstruction, involving large obstacles, is uncommon and currently not supported by our system.

\parab{Interference from pouch.} Although chirp signals can resist interference and noise, they fail when the interference spans the entire frequency range in low SNR cases as shown in Fig.~\ref{fig:spectrogram}.
In our experiment, the soft pouch generated such interference when it deforms due to movement. A practical solution is to use a hard case instead of a soft pouch with extra attention brought by the hard case. Alternatively, filling the pouch with more air can help resist shape changes and reduce interference.

\parab{End-to-end latency.}
Based on the latency breakdown discussed in \S\ref{sec:eval_system_metrics}, several methods can be implemented to reduce the end-to-end delay. First, image encoding and decoding currently run on CPU which is time-consuming but they can be migrated to on-device GPUs to speed up the inference with Vulkan backend~\cite{vulkan_2024}.
Second, we can hide most of the processing latency via pipelining and only the transmission latency becomes the bottleneck.

\parab{Preamble detection.}
Current preamble design and detection algorithm is vulnerable to interference and can be further improved. In the future, more advanced techniques such as neural network-based preamble design and detection~\cite{yang80neural}, can be leveraged to improve the packet detection reliability.

\section{Related Work}
\label{sec:related_work}

\subsection{Underwater Communication}
\noindent\textbf{Mobile devices.} AquaApp~\cite{chen2022underwater} is the first work that enables underwater text message transmission between two mobile phones. They further extend the system for 3D localization underwater~\cite{chen2023underwater}.
AquaHelper~\cite{yang2023aquahelper,yang80neural} focuses on designing robust signal detection algorithms for SOS message transmission.
The information transmitted by their systems is very limited.
Although some work in~\cite{liu2021uqcom,liu2023uqrcom} attempts to communicate via QR codes on mobile devices. Successfully scanning the QR code heavily depends on the underwater visibility and the distance between the sender and receiver. Instead, \sysname is capable of transmitting images on mobile devices.

\parab{Customized hardware.}
Researchers in \cite{afzal2022battery,mobicom24_seascan} transmit an RGB image via a Piezo-acoustic backscatter~\cite{jang2019underwater,rademacher2022enabling}. Due to its low data rate, it takes hours to send a raw RBG data.
Amphilight~\cite{carver2021amphilight} and Shrimp~\cite{lin2021shrimp} employ optical techniques for underwater transmission, utilizing customized LED lights and laser diodes, respectively. However, the optics-based communication system is mainly limited by the transmission distance and the light condition underwater.
Unlike others, \sysname focuses on enabling underwater communication on commercial off-the-shelf mobile phones, making the technology more accessible and ubiquitous.

\parab{Chirp-based acoustic systems.}
Chirp-based transmission protocol is widely adopted for underwater communication~\cite{xie2024icc,jia2022two,steinmetz2018practical,steinmetz2022taking,restuccia2017isonar,tonolini2018networking}. They mainly focus on the optimization of the transmission protocols such as rate adaptation, modulation designs, multi-path elimination. Their designed protocols still are for a small packet and require larger transmission power and bandwidth which cannot be directly applied for image transmission on mobile devices.
\sysname optimizes the chirp-based protocol tailored for robust image transmission on commercial mobile devices.

\parab{Underwater image transmission.}
Researchers in \cite{anjum2022deep,anjum2022acoustic} utilize the traditional CNN-based image compression approach to reduce image size. However, as discussed in \S\ref{sec:mov_challenges}, the images compressed by those approaches do not meet the required size by mobile devices and are not robust to transmission errors. \sysname utilizes the state-of-the-art generative compression models which are bandwidth-efficient and error-resilient.

\subsection{Generative Compression}
Generative compression has attracted significant attention \cite{grassucci2023generative,xie2021tsp,dai2023wc,anjum2022deep,le2023arxiv,li2024reparolossresilientgenerativecodec,NEURIPS2019_f1ea154c} and has been successfully applied to image/video codecs and wireless communication, where it reduces communication overhead without compromising the quality of the received images or videos.
\sysname is the first system to apply generative compression to underwater communications, and it also extends this capability to other limited-bandwidth or error-prone scenarios---such as satellite networks and rural communication.

Image tokenization~\cite{yan2024elastictokadaptivetokenizationimage,shen2025catcontentadaptiveimagetokenization}, as a way for generative compression, learns a compact image representation for various downstream tasks such as classification, reconstruction, and generation.
Since the introduction of neural discrete representation~\cite{van2017neural}, numerous methods~\cite{esser2020taming, yu2024image, yu2023language, yu2021vector} have been developed. Among these, VQGAN~\cite{esser2020taming} and TiTok~\cite{yu2024image} were selected for our system due to their balance between compressed image size and decoded image quality.

\section{Conclusion}
\label{sec:conclusion}

We present \sysname, the first underwater acoustic system that enables
image transmission on mobile devices.
By combining generative image compression with reliability-enhancing techniques at the PHY layer,
\sysname transmits 256$\times$256 color images with low latency
across various underwater conditions,
preserving high perceptual quality even under stringent bandwidth limitations
and severe transmission errors.
By extending underwater communication from text messaging to image
transmission, \sysname contributes to the ongoing development of underwater
technologies.

\parab{Ethics.} This work does not raise any ethical issues.

\bibliographystyle{ACM-Reference-Format}
\bibliography{ref}

\end{document}